\documentclass[11pt]{article}
\usepackage{graphicx,latexsym}
\usepackage{amsmath}
\usepackage{epstopdf}
\usepackage{color}
\usepackage{subfigure}
\usepackage{makecell}
\newcommand{\xvec}{{\bf x}}

\newcommand{\lxor}{\oplus}
\newcommand{\lnon}{\overline}

\newcommand{\qed}{$\Box$}

\newcommand{\minbn}{{\bf MinContBN}}

\newtheorem{proposition}{Proposition}
\newtheorem{lemma}{Lemma}
\newtheorem{theorem}{Theorem}

\newtheorem{example}{Example}
\newtheorem{remark}{Remark}

\begin{document}

\title{On the Number of Control Nodes in Boolean Networks with Degree Constraints}

\author{Liangjie Sun$^1$, Wai-Ki Ching$^2$, Tatsuya Akutsu$^1$$^*$}
\date{
   $^1$Bioinformatics Center, Institute for Chemical Research, Kyoto University, Kyoto 611-0011, Japan\\
   $^2$Department of Mathematics, The University of Hong Kong, Pokfulam Road, Hong Kong\\
   $^*$takutsu@kuicr.kyoto-u.ac.jp}

\maketitle
\begin{abstract}
This paper studies the minimum control node set problem for Boolean networks (BNs) with degree constraints. The main contribution is to derive the nontrivial lower and upper bounds on the size of the minimum control node set through combinatorial analysis of four types of BNs (i.e., $k$-$k$-XOR-BNs, simple $k$-$k$-AND-BNs, $k$-$k$-AND-BNs with negation and $k$-$k$-NC-BNs, where the $k$-$k$-AND-BN with negation is an extension of the simple $k$-$k$-AND-BN that considers the occurrence of negation and NC means nested canalyzing). More specifically, four bounds for the size of the minimum control node set: general lower bound, best case upper bound, worst case lower bound, and general upper bound are studied.
%, where the general lower bound is a value that is not less than the size of the control node set for any BN, the general upper bound is the maximum value of the size of the minimum control node set for any BN, while the best case upper bound (resp., the worst case lower bound) is the minimum (resp., maximum) value currently found, which is obtained from some BN.
By dividing nodes into three disjoint sets, extending the time to reach the target state, and utilizing necessary conditions for controllability, these bounds are obtained, and further meaningful results and phenomena are discovered. Notably, all of the above results involving the AND function also apply to the OR function.

{\bf Keywords:} Boolean networks, controllability, minimum control node set.
\end{abstract}
\section{Introduction}
Revealing the dynamics of the macromolecular interaction network behind cellular functions is one of the significant tasks of systems biology. The control of the internal state of cells to transition from an initial state to an ultimate target state has been widely discussed and studied, due to its wide range of application prospects, such as finding new therapeutic targets for diseases to transform the disease cell state into a normal state \cite{barabasi11,pun23}, stem cell reprogramming to systematically control the reprogramming or transdifferentiation of differentiated cells \cite{takahashi06,young11}, etc. However, since manipulating the activity of even a single intracellular component is a long, difficult, and expensive experimental task, it is critical to minimize the number of nodes that need to be controlled.

In recent years, the control theory framework has been extended to Boolean network (BN) models \cite{akutsu12,Toyoda24,yu24,ni24,liy24,wu24,wang25}. BN, proposed by Kauffman \cite{kauffman69} and Thomas \cite{thomas73} as a prototype model for gene regulatory networks, is a logical dynamical system composed of a large number of highly interconnected nodes whose states (0 or 1) are updated by Boolean functions of other nodes and/or themselves. A number of achievements have been made in finding the optimal driver set with specific attractors for a given BN. For example, a useful concept called the control kernel was introduced by Kim et al. \cite{kim13}, which is a minimal set of nodes that need to be controlled to guide the state of the BN to converge to the desired attractor regardless of the initial state of the BN, and a general algorithm was further developed to identify the control kernel based on the genetic algorithm and the attractor landscape of the network. Subsequently, using BN models, Za{\~n}udo and Albert \cite{zanudo15} proposed a network control method that integrates network structure and functional information to find partial fixed points (called stable motifs) of network dynamics that steer the dynamics to the desired attractor. Hou et al. \cite{hou19} mathematically proved that under reasonable assumptions, a small number of driver nodes are sufficient to control a BN to the target attractor.
Borriello and Daniels \cite{borriello21} then analyzed the control kernels of 49 biological network models and found that the average number of nodes that must be controlled is logarithmic with the number of original attractors. Later, a feasible approximation method was found by Parmer et al. \cite{parmer22} to determine the nodes that control the dynamics of the desired attractor based on the idea of maximizing the influence of the propagation process in social networks.

On the other hand, many studies focus on finding the minimal set of key variables in the BN and thus controlling them to make the system controllable. Here, controllability \cite{cheng09,laschov12,lir15,weiss18,li19,li24} means that for any pair of states $\xvec^0,\xvec^T\in\{0,1\}^{n}$, there exists a control sequence steering the state from $\xvec(0) = \xvec^0$ to $\xvec(t) = \xvec^T$ for some integer $t$. For example, by using graph theory method, the minimal controllability of conjunctive BN (an important subclass of BN) was explored and a sufficient and necessary condition for controllability of conjunctive BN was derived in \cite{weiss18}. Afterwards, the minimal strong structural controllability problem of BNs was studied in \cite{zhu22}, and then an effective method for identifying the minimum set of controlled nodes to achieve global stabilization of large-scale BNs was presented in \cite{zhu23}.

However, the above-mentioned studies have not further analyzed the minimum number of control nodes required to achieve controllability of BNs from a theoretical perspective, which is a meaningful and important topic. In this paper, we focus on the minimum control node set problem for BNs with degree constraints, and mainly consider four types of BNs, where the indegree and outdegree of each node are both $k$: (i) $k$-$k$-XOR-BN, where each Boolean function only contains XOR operations, (ii) simple $k$-$k$-AND-BN (resp., simple $k$-$k$-OR-BN), where all Boolean functions are simple AND functions (resp., OR functions) without negations, (iii) $k$-$k$-AND-BN with negation (resp., $k$-$k$-OR-BN with negation), that is an extension of simple $k$-$k$-AND-BN (resp., simple $k$-$k$-OR-BN) considering the occurrence of literal $\lnon{x}$, and (iv) $k$-$k$-NC-BN, where each Boolean function is a nested canalyzing function \cite{Kauffman2004,akutsu11}. Note that considering the degree constraint is reasonable because many biologically relevant Boolean functions have a small number of input variables \cite{harris2002}.

Specifically, we define four concepts for the size of the minimum control node set: general lower bound, general upper bound, worst case lower bound, and best case upper bound (see Table \ref{table} for details), and separately consider the nontrivial upper and lower bounds on the size of the minimum control node set required for each type of BNs.
\begin{table*}
  \centering
  \caption{The meaning of each bound, where $b$ is the corresponding bound.}
  \resizebox{\linewidth}{!}{
  \label{table}
  \begin{tabular}{|c|p{12cm}|}
   \hline
   General lower bound & For any BN, the number of control nodes is not smaller than $b$. \\ \hline
   Best case upper bound & For some BN, the minimum number of control nodes is at most $b$, which is currently the smallest. \\ \hline
   Worst case lower bound & For some BN, the number of control nodes is not smaller than $b$, which is currently the largest. \\ \hline
   General upper bound & For any BN, the minimum number of control nodes is at most $b$. \\ \hline
  \end{tabular}}
\end{table*}
Generally speaking, the following relationship holds: general lower bound $\leq$ best case upper bound $<$ worst case lower bound $\leq$ general upper bound. If general lower bound $=$ best case upper bound (resp., worst case
lower bound $=$ general upper bound), these bounds are called tight or optimal \cite{sun24}. In this paper, $n$ represents the total number of nodes in a BN, and it is clear that $n$ is a trivial general upper bound. If the worst case lower bound is also $n$, then it is tight. The main results are summarized in Table \ref{table2} and
their meanings are discussed in Section \ref{section-6}.
\begin{table*}
  \centering
  \caption{Summary of results, where $n$ represents the total number of nodes in a BN, $t^{*}$ is the minimum time that any initial state $\xvec^0$ can reach any target state $\xvec^T$, $\beta(k)=\max\left\{\frac{k-l}{2k-l}n+\frac{1}{2k-l}\left(\sum_{j=1}^{l-2}j\cdot|M_{j}|\right)+\frac{l-1}{2k-l}\left(\sum_{j=l-1}^{\lfloor\frac{k}{2}\rfloor}|M_{j}|\right),\;l=1,2,\ldots,\lfloor\frac{k}{2}\rfloor+1\right\}$, $M_{j}$ is the set of nodes where the literal $x_{i}$ or $\lnon{x_{i}}$ appears $j$ times in the $k$-$k$-AND-BN with negation, and for any $k_{1}<k_{2}$, $\sum_{j=k_{2}}^{k_{1}}|M_{j}|=0$. }
\label{table2}
\resizebox{\linewidth}{!}{
\begin{tabular}{|l|c|c|c|}
   \hline
  & Type & Control node set  & Characteristic\\ \hline
  Proposition \ref{prop:xor-lower}& 2-2-XOR-BNs ($n$ is even) & $\frac{1}{2}n$ & Best case upper bound\\ \hline
  Theorem \ref{thm:xor-upper} & 2-2-XOR-BNs ($n=3m$ and $m>1$ is even) & $\frac{5}{6}n$ & General upper bound\\ \hline
  Theorem \ref{thm:xor-time} & 2-2-XOR-BNs & \makecell[c]{$\frac{n}{t^{*}}$\\$n-t^{*}+1$} & \makecell[c]{General lower bound\\General upper bound}\\ \hline
  Remark \ref{remark2}& $k$-$k$-XOR-BNs & \makecell[c]{$\frac{n}{t^{*}}$\\$n-t^{*}+1$} & \makecell[c]{General lower bound\\General upper bound}\\ \hline
  Theorem \ref{thm:4-4-xor} & $k$-$k$-XOR-BNs ($n=(k+1)m$ and $k\geq3$ is odd) & $k+1$ & Best case upper bound\\ \hline
  Theorem \ref{theorem4} & $k$-$k$-XOR-BNs ($n=[k(k-1)+1]m$ and $(n \mod k)=0$) & $\left[1- \frac{k-1}{k(k^{2}-k+1)}\right]n$ & General upper bound\\ \hline
  Proposition \ref{proposition4} & simple 2-2-AND-BN ($n$ is even) & $n$ & Worst case lower bound\\ \hline
  Proposition \ref{proposition6} & simple 2-2-AND-BN ($n=3m+1$) & $\frac{n-1}{3}+1$ & Best case upper bound\\ \hline
  Theorem \ref{thm:2-2-and} & simple 2-2-AND-BN & $\frac{1}{3}n$ & General lower bound\\ \hline
  Proposition \ref{proposition7} & simple $k$-$k$-AND-BN ($n=(2k-1)m+1$ and $m>1$) & $\frac{k-1}{2k-1}(n-1) + 1$ & Best case upper bound\\ \hline
  Theorem \ref{thm:k-k-AND} & simple $k$-$k$-AND-BN & $\frac{k-1}{2k-1}n$ & General lower bound\\ \hline
  Theorem \ref{thm:2-2-AND-negation} & 2-2-AND-BN with negation & $\max\left(\frac{1}{3}n,\frac{1}{2}|M_{1}|\right)$ & General lower bound\\ \hline
  Theorem \ref{thm:k-k-AND-negation} & $k$-$k$-AND-BN with negation & $\beta(k)$ & General lower bound\\ \hline
  Proposition \ref{prop:nc} & $k$-$k$-NC-BN & $\frac{n}{k}$ & Best case upper bound\\ \hline
\end{tabular}}
\end{table*}
\section{Problem Definition}
Consider the following synchronous BN:
\begin{eqnarray*}
x_i(t+1) & = & f_i(x_1(t),\ldots,x_n(t)),\quad \forall i \in [1,n],
\end{eqnarray*}
where every $x_{i}(t)$ takes values in $\{0,1\}$ and $f_{i}:\{0,1\}^{n}\rightarrow\{0,1\}$ is a Boolean function. Here,
we use $x_i$ to denote the node corresponding to $x_i(t)$
and $V=\{x_1,\ldots,x_n\}$ to denote the set of nodes in the BN.
Let $\xvec(t)=[x_1(t),\ldots,x_n(t)]$,
and use $N(V,F)$ to denote this BN, where $F = \{f_1,\ldots,f_n\}$.
For each state vector $\xvec$,
$(\xvec)_i$ denotes the state of the $i$th element of $\xvec$.
Further, $F(\xvec)$ is used to denote the state vector
at the next time step (i.e., $\xvec(t+1) = F(\xvec(t))$), and
$G(V,E)$ is used to denote the underlying directed graph,
that is, $(x_j,x_i) \in E$ if and only if $x_j$ is an input node of $x_i$. In addition, if there exists an edge from $x_{j}$ to $x_{i}$, then $x_{j}$ is called an incoming node of $x_{i}$, and $x_{i}$ is called an outgoing node of $x_{j}$. The indegree and outdegree of a node $x_{i}$ are defined as the numbers of edges directed toward $x_{i}$ and from $x_{i}$ to other nodes, respectively.

We choose a subset of nodes $U \subseteq V$ as a set of control nodes
and assume that
the dynamics of the corresponding BN with a set of control nodes $U$
can be given by
\begin{eqnarray*}
x_i(t+1) & = & f_i(x_1(t),\ldots,x_n(t)),\quad \forall x_i \notin U,\\
x_i(t+1) & = & u_i(t) \lxor f_i(x_1(t),\ldots,x_n(t)),\quad \forall x_i \in U,
\end{eqnarray*}
where $u_i(t)$ can take an arbitrary value in $\{0,1\}$ and
$x \lxor y$ denotes the exclusive OR of two Boolean variables $x$ and $y$.

Then, we consider the following problem.

\bigskip

\noindent
{\bf Problem: MinContBN}\\
Given a BN $N(V,F)$,
find a minimum cardinality set $U \subseteq V$ such that
for any initial state $\xvec^0$ and any target state $\xvec^T$,
there exist control signals $u_i(t)$ for all $x_i \in U$ that drive
$N(V,F)$ from $\xvec(0) = \xvec^0$ to $\xvec(t) = \xvec^T$ for some $t$.

\bigskip

$U$ in {\minbn} is referred to as the \emph{minimum control node set},
and (not necessarily minimum) $U$ satisfying the conditions above is
referred to as a \emph{control node set}.

\section{Boolean networks consisting of XOR functions}
In this section, we consider the case where either one of the following functions is
assigned to each node:
\begin{eqnarray*}
x_i(t+1) & = & x_{i_1}(t) \lxor x_{i_2}(t),\\
x_i(t+1) & = & \lnon{x_{i_1}(t) \lxor x_{i_2}(t)}.
\end{eqnarray*}
This BN is referred to as 2-XOR-BN.

\begin{proposition}
For each even $n$, there exists a 2-XOR-BN with a control node set of size 2.
\end{proposition}
(Proof)
For each of $i=3,\ldots,n$,
we let
\begin{eqnarray*}
x_i(t+1) & = & x_1(t) \lxor x_{i-1}(t),
\end{eqnarray*}
and we let
\begin{eqnarray*}
x_1(t+1) & = & x_1(t) \lxor x_2(t),\\
x_2(t+1) & = & x_1(t) \lxor x_2(t),
\end{eqnarray*}
where any XOR function can be assigned to each of
$x_1$ and $x_2$.

Then, let $U=\{x_1,x_2\}$, and it is straightforward to see that $U$ satisfies the conditions for the
control node set.
\qed

\bigskip

In this case, the outdegree of $x_1$ is quite large.
Hence, it is reasonable to ask what happens if the outdegree of each node is two.
We use 2-2-XOR-BN to denote such a BN.

\begin{proposition}\label{prop:xor-lower}
For each even $n$, there exists a 2-2-XOR-BN for which the size of the minimum control node set
is $\frac{n}{2}$.
\end{proposition}
(Proof)
For each $i=1,\ldots,\frac{n}{2}$, we let
\begin{eqnarray*}
x_{2i-1}(t+1) & = & x_{2i-1}(t) \lxor x_{2i}(t),\\
x_{2i}(t+1) & = & x_{2i-1}(t) \lxor x_{2i}(t).
\end{eqnarray*}
Clearly, for each pair $(x_{2i-1},x_{2i})$,
it is necessary and sufficient to
let either $x_{2i-1}$ or $x_{2i}$ as a control node.
\qed

\bigskip

Next, we derive a general upper bound on the size of
the control node set for 2-2-XOR-BNs.

Let $\Gamma^+(x_i)$ denote
the set of outgoing nodes (i.e.,
$\Gamma^+(x_i) = \{x_j \mid (x_i,x_j) \in E\}$).
Similarly, $\Gamma^-(x_i)$ denotes the set of incoming nodes.
Note that $|\Gamma^+(x_i)| = |\Gamma^-(x_i)| = 2$ holds for each $x_i$
in any 2-2-XOR-BN.

The following lemma is simple but plays an important role.

\begin{lemma}\label{lemma1}
For any 2-2-XOR-BN with $n=3m$ for some integer $m$,
there exists $V_1 \subseteq V$ with $|V_1| = {\frac 1 3}n = m$
such that $\Gamma^{+}(x_i) \cap \Gamma^+(x_j) = \emptyset$
holds for any $x_i \neq x_j$ in $V_1$.
\end{lemma}
(Proof)
We use the following greedy procedure to find a required $V_1$,
\begin{description}
\item [STEP 1] Let $V_1 \leftarrow \emptyset$ and $S \leftarrow V$.
\item [STEP 2] Repeat STEP 3 and STEP 4 $m$ times.
\item [STEP 3] Select an arbitrary element $x_i \in S$ and let
$V_1 \leftarrow V_1 \cup \{x_i\}$.
\item [STEP 4] Let $S \leftarrow S \setminus (\{x_i\} \cup
\{x_j \in S \mid \Gamma^+(x_j) \cap \Gamma^+(x_i) \neq \emptyset \})$.
\end{description}
Note that there exist at most two such $x_j$s at STEP 4 because of
the constraints on indegree and outdegree.
Consequently, the size of $S$ decreases by at most three for each iteration.
Therefore, we can repeat STEP 3 and STEP 4 $m$ times and thus
can get the required $V_1$.
\qed

\bigskip

We partition $V$ into three disjoint sets $V_1$, $V_2$, and $V_3$, where
$V_2 = \{x_j \mid (x_i,x_j) \in E ~\land~ x_i \in V_1 ~\land~ x_j \notin V_1 \}$ and
$V_3 = V \setminus (V_1 \cup V_2)$.
Then, it is straightforward to see that $|V_1|=m$, $|V_2| \geq m$,
and $|V_3| \leq m$ always hold.

\begin{theorem}\label{thm:xor-upper}
Suppose that $n=3m$ for some even number $m > 1$.
Then, for any 2-2-XOR-BN,
there exists a control node set of size ${\frac {5} {6}}n$.
\end{theorem}
(Proof)
Let $N(V,F)$ be a given 2-2-XOR-BN.
We assume without loss of generality $V_1 = \{x_1,\ldots,x_m\}$
(by appropriately permuting indices of $x_i$s).

We partition $V_1$ into three disjoint sets
$V_1^0$, $V_1^1$, and $V_1^2$, where
\begin{eqnarray*}
V_1^0 & = & \{x_i \in V_1 \mid |\Gamma^+(x_i) \cap V_1|=0\},\\
V_1^1 & = & \{x_i \in V_1 \mid |\Gamma^+(x_i) \cap V_1|=1\},\\
V_1^2 & = & \{x_i \in V_1 \mid |\Gamma^+(x_i) \cap V_1|=2\}.
\end{eqnarray*}
For each node $x_i \in V_1^0 \cup V_1^1$,
we let $\phi(i)=j_2$ where $\Gamma^+(x_i)=\{x_{j_1},x_{j_2}\}$ and $j_1 < j_2$.
Clearly, $x_{\phi(i)} \in V_2$ holds.
$\phi(i)$ is introduced so that each $x_{\phi(i)}$ takes the desired value
at $t=2$ by giving an appropriate control to $x_i$ at $t=1$.

We define a set of node $W$ by
$$
W = \{ x_{j_1} \mid \Gamma^+(x_i)=\{x_{j_1},x_{j_2}\} ~\land~
x_i \in V_1^0 ~\land~ j_1 < j_2\}.
$$
Since $\bigcup_{x_i \in V_1^0} \Gamma^+(x_i) \subseteq V_2$,
$W \subseteq V_2$ and $|W| \leq {\frac 1 2} |V_2|$ hold.
Note also that $x_{\phi(i)} \notin W$ for any $x_i \in V_1^0 \cup V_1^1$.

Then, let $U = V_1 \cup W \cup V_3$ be the set of control nodes.
From $|V_2| \geq m$, $|V_3|=n-|V_1|-|V_2|$, and $|W| \leq {\frac 1 2} |V_2|$,
we have
\begin{eqnarray*}
|U| & = & |V_1| + |W| + |V_3|\\
& = & |V_1| + |W| + n - |V_1| - |V_2|\\
& = & n - |V_2| + |W|\\
& \leq & n - {\frac 1 2}|V_2| \\
& \leq & n - {\frac m 2} \\
&=& {\frac 5 6} n.
\end{eqnarray*}

Next, we show that $U$ satisfies the conditions of control node set.
Let $\xvec^0$ and $\xvec^T$ be specified initial and target state vectors,
respectively.
Let $\xvec^1 = F(\xvec^0)$ and $\xvec^2=F(\xvec^1)$ (without any control).
Then, controls are given as below.

At $t=0$,
for each node $x_i \in V_1^0 \cup V_1^1$ with
$(\xvec^2)_{\phi(i)} \neq (\xvec^T)_{\phi(i)}$,
we let $u_i(0) = 1$.
For any other control node $x_i$,
we let $u_i(0) = 0$.
Let the resulting state vectors at $t=1$ and $t=2$
be $\xvec^3$ and $\xvec^4$, respectively.
Then, $(\xvec^4)_{\phi(i)} = (\xvec^T)_{\phi(i)}$ holds for
all $\phi(i)$.

At $t=1$,
for each node $x_i \in U$,
we let $u_i(1)=1$ if $(\xvec^4)_i \neq (\xvec^T)_i$,
otherwise we let $u_i(1)=0$.
Then, the resulting state vector $\xvec^5$ (at $t=2$) satisfies
$\xvec^5 = \xvec^T$.
\qed

\bigskip

\begin{figure}[th]
\begin{center}
\includegraphics[width=14cm]{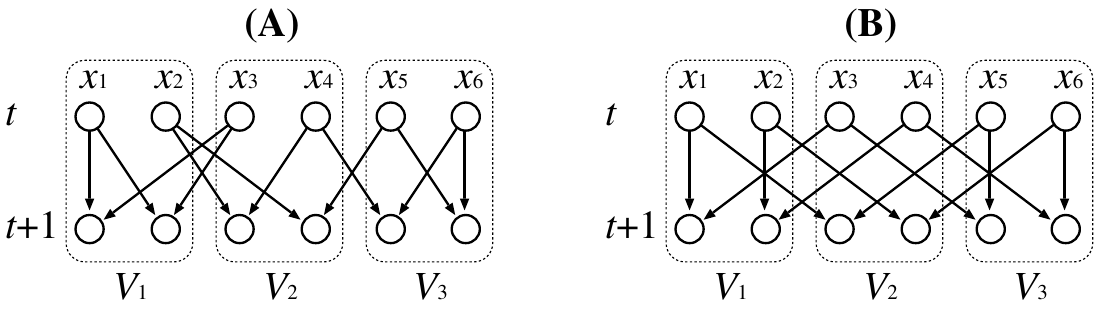}
\caption{2-2-XOR-BNs for Examples \ref{ex:xor1} and \ref{ex:xor2}.}
\label{fig:xor}
\end{center}
\end{figure}

\begin{example}
\label{ex:xor1}
Consider a BN given in Fig.~\ref{fig:xor}(A),
where all Boolean functions are of type $x \lxor y$.
In this case, $W=\{x_3\}$, $U=\{x_1,x_2,x_3,x_5,x_6\}$,
$V_1^0=\{x_2\}$,
$V_1^1=\emptyset$,
$V_1^2=\{x_1\}$,
and $\phi(2)=4$.

Suppose $\xvec^0=[0,0,0,0,0,0]$ and $\xvec^T=[1,1,1,1,1,1]$.
Then, we have $\xvec^2=[0,0,0,0,0,0]$ (at $t=2$).

At $t=0$, we let $u_2(0)=1$ and $u_{1}(0)=u_{3}(0)=u_{5}(0)=u_{6}(0)=0$.
At $t=1$, we let $u_1(1)=u_2(1)=u_5(1)=u_6(1)=1$ and $u_3(1)=0$.
Then, we have the following.
\begin{center}
\begin{tabular}{l|llllll}
\hline
$\xvec^0$ ($t=0$) & 0 & 0 & 0 & 0 & 0 & 0\\
$\xvec^1$ ($t=1$) & 0 & 0 & 0 & 0 & 0 & 0\\
$\xvec^2$ ($t=2$) & 0 & 0 & 0 & 0 & 0 & 0\\
\hline
$\xvec^3$ ($t=1$) & 0 & 1 & 0 & 0 & 0 & 0\\
$\xvec^4$ ($t=2$) & 0 & 0 & 1 & 1 & 0 & 0\\
$\xvec^5$ ($t=2$) & 1 & 1 & 1 & 1 & 1 & 1\\
\hline
\end{tabular}
\end{center}
where
$\xvec^1$ and $\xvec^2$ denote the resulting states at $t=1$ and $t=2$ without any control, $\xvec^3$ represents the state at $t=1$ when the control signals $u_{2}(0)=1$ and $u_{1}(0)=u_{3}(0)=u_{5}(0)=u_{6}(0)=0$ are applied at $t=0$,
$\xvec^4$ denotes the state at $t=2$ under the same control condition at $t=0$ without applying any control at $t=1$,
and $\xvec^5$ represents the state at $t=2$ when the control signals $u_{2}(0)=1,u_{1}(0)=u_{3}(0)=u_{5}(0)=u_{6}(0)=0$ are applied at $t=0$ and $u_1(1)=u_2(1)=u_5(1)=u_6(1)=1,u_3(1)=0$ are applied at $t=1$.
\end{example}

\begin{example}
\label{ex:xor2}
Consider a BN given in Fig.~\ref{fig:xor}(B),
where all Boolean functions are of type $x \lxor y$.
In this case, $W=\emptyset$, $U=\{x_1,x_2,x_5,x_6\}$,
$V_1^0=\emptyset$,
$V_1^1=\{x_1,x_2\}$,
$V_1^2=\emptyset$,
and $\phi(1)=3$ and $\phi(2)=4$.

Suppose $\xvec^0=[0,0,0,0,0,0]$ and $\xvec^T=[1,1,1,1,1,1]$.
Then, we have $\xvec^2=[0,0,0,0,0,0]$ (at $t=2$).

At $t=0$, we let $u_1(0)=u_2(0)=1$ and $u_5(0)=u_6(0)=0$.
At $t=1$, we let $u_1(1)=u_2(1)=0$ and $u_5(1)=u_6(1)=1$.
Then, we have the following.
\begin{center}
\begin{tabular}{l|llllll}
\hline
$\xvec^0$ ($t=0$) & 0 & 0 & 0 & 0 & 0 & 0\\
$\xvec^1$ ($t=1$) & 0 & 0 & 0 & 0 & 0 & 0\\
$\xvec^2$ ($t=2$) & 0 & 0 & 0 & 0 & 0 & 0\\
\hline
$\xvec^3$ ($t=1$) & 1 & 1 & 0 & 0 & 0 & 0\\
$\xvec^4$ ($t=2$) & 1 & 1 & 1 & 1 & 0 & 0\\
$\xvec^5$ ($t=2$) & 1 & 1 & 1 & 1 & 1 & 1\\
\hline
\end{tabular}
\end{center}
where $\xvec^1,\ldots,\xvec^5$ are defined as in Example \ref{ex:xor1}, except that the control signals at $t=0$ and $t=1$ differ.

Next, suppose $\xvec^0=[1,0,0,1,1,0]$ and $\xvec^T=[0,1,0,1,1,0]$.
Then, we have $\xvec^2=[1,1,0,0,1,1]$ (at $t=2$).

At $t=0$, we let $u_2(0)=1$ and $u_1(1)=u_5(1)=u_6(1)=0$.
At $t=1$, we let $u_1(1)=u_2(1)=u_6(1)=1$ and $u_5(1)=0$.
Then, we have the following.
\begin{center}
\begin{tabular}{l|llllll}
\hline
$\xvec^0$ ($t=0$) & 1 & 0 & 0 & 1 & 1 & 0\\
$\xvec^1$ ($t=1$) & 1 & 1 & 0 & 0 & 1 & 1\\
$\xvec^2$ ($t=2$) & 1 & 1 & 0 & 0 & 1 & 1\\
\hline
$\xvec^3$ ($t=1$) & 1 & 0 & 0 & 0 & 1 & 1\\
$\xvec^4$ ($t=2$) & 1 & 0 & 0 & 1 & 1 & 1\\
$\xvec^5$ ($t=2$) & 0 & 1 & 0 & 1 & 1 & 0\\
\hline
\end{tabular}
\end{center}
where $\xvec^1,\ldots,\xvec^5$ are defined as in Example \ref{ex:xor1}, except that the control signals at $t=0$ and $t=1$ differ.
\end{example}

For the 2-2-XOR-BN in Proposition \ref{prop:xor-lower}, we have $V_{1}=\{x_{2i-1}\}$, $V_{2}=\{x_{2i}\}$ and $V_{3}=\emptyset$, where $i=1,\ldots,\frac{n}{2}$. According to Theorem \ref{thm:xor-upper}, we note that $W=\emptyset$, $U=\{x_{2i-1}\}$, $V_{1}^{0}=\emptyset$, $V_{1}^{1}=\{x_{2i-1}\}$, $V_{1}^{2}=\emptyset$, and $\phi(2i-1)=2i$, where $i=1,\ldots,\frac{n}{2}$ and $|U|=\frac{n}{2}$. Therefore, the 2-2-XOR-BN in Proposition \ref{prop:xor-lower} is a special case.

In order to shorten the gap between the lower bound in Proposition \ref{prop:xor-lower} and
the upper bound in Theorem \ref{thm:xor-upper}, we have the following results.

\begin{proposition}\label{xor-special1}
Suppose that $n=3m$ for some number $m > 1$.
Then, for any 2-2-XOR-BN, where $V_{1}^{0}=\emptyset$,
there exists a control node set of size ${\frac {2} {3}}n$.
\end{proposition}
(Proof)
First, for any 2-2-XOR-BN, where $V_{1}^{0}=\emptyset$, we have $V_{1}^{2}=\emptyset$ and $|V_{1}^{1}|=|V_{2}|=m$. If $V_{1}^{0}=\emptyset$, then $|\Gamma^+(x_i)\cap V_{1}|\geq1$ for any $x_{i}\in V_{1}$. Since $|V_{1}|=m$ and $\Gamma^+(x_i)\cap\Gamma^+(x_j)=\emptyset$ for any $x_{i}\neq x_{j}$ in $V_{1}$, we have $|\Gamma^+(x_i)\cap V_{1}|=1$ for any $x_{i}\in V_{1}$, which means that $V_{1}^{2}=\emptyset$ and $V_{1}=V_{1}^{1}$. And when $|V_{1}|=|V_{1}^{1}|=m$, we have $|V_{2}|=|V_{1}^{1}|=m$ and $|V_{3}|=|V|-|V_{1}^{1}|-|V_{2}|=m$.

In this case, we note that $W=\emptyset$ and $U=V_{1}\cup V_{3}$, where $|U|=2m={\frac {2} {3}}n$.
\qed

\bigskip

\begin{remark}
According to Proposition \ref{xor-special1}, we note that for any 2-2-XOR-BN, where $V_{1}^{0}=\emptyset$ and $n=2m$, if $V_{3}=\emptyset$, i.e., $\Gamma^{+}(x_{j})\subseteq V_{1}\cup V_{2}$ for any $x_{j}\in V_{2}$, then $U=V_{1}^{1}$, where $|U|=m=\frac{n}{2}$.
\end{remark}

It is worth noting that, based on the above results, for any initial state $\xvec^0$, by giving appropriate control signals $u_{i}(0)$ and $u_{i}(1)$, where $x_{i}\in U$, the target state $\xvec^T$ can be reached at $t=2$. Next, we consider whether it is possible to further reduce the size of the control node set. We first present the following example.

\begin{example}\label{ex:xor3}
Consider a BN given in Fig. \ref{fig:xor2a}, where all Boolean functions are of type $x \lxor y$.
\begin{figure}[htb]
\centering
\subfigure[]{\includegraphics[width=0.35\textwidth]{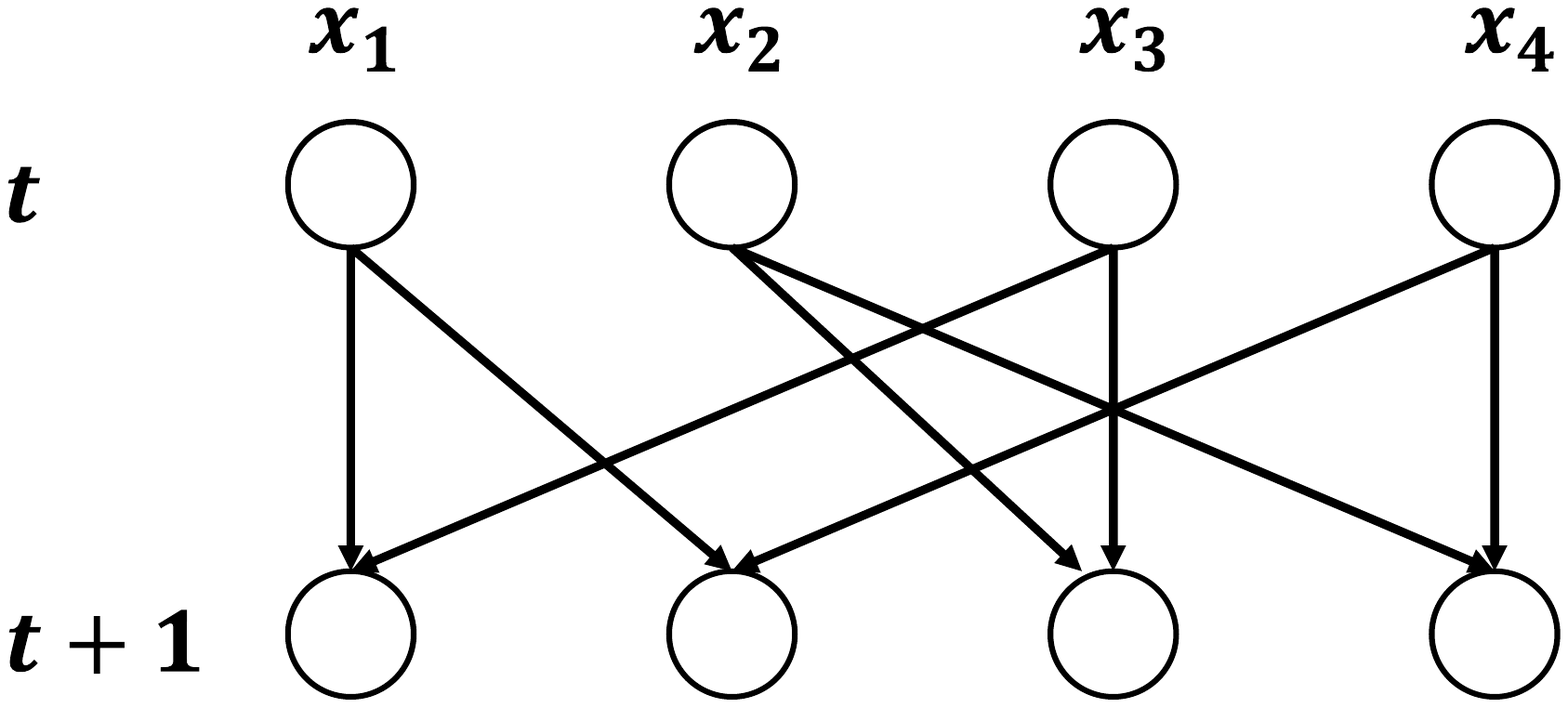}\label{fig:xor2a}}
\hfil
\subfigure[]{\includegraphics[width=0.55\textwidth]{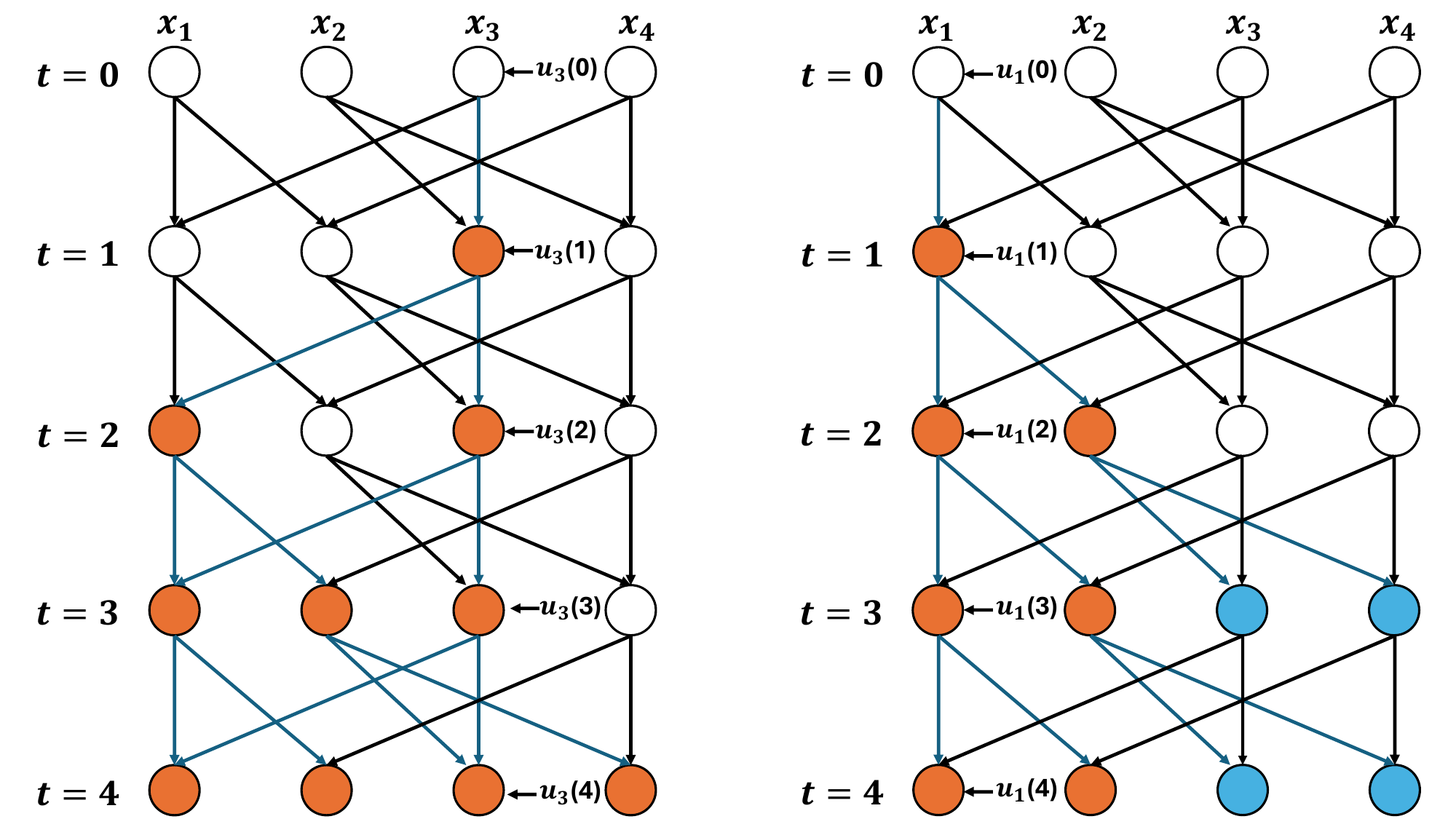}\label{fig:xor2b}}
\caption{(a) A 2-2-XOR-BN for Example \ref{ex:xor3}. (b) The value of nodes under different controls.}
\label{fig:xor2}
\end{figure}

In this case, let $U=\{x_{3}\}$. Based on the Table \ref{table1}, it is seen that by giving appropriate control signals $u_{3}(t),~t\in[0,3]$, we can drive this BN from any initial stats $\xvec^0$ to any target states $\xvec^T$ at $t=4$. Specifically, according to Table \ref{table1}, we find that (Fig. \ref{fig:xor2b})
\begin{itemize}
  \item $x_{3}$ can take any 2 values (0 or 1) from $t=1$;
  \item $(x_{1},x_{3})$ can take any 4 values (00, 10, 01 or 11) from $t=2$;
  \item $(x_{1},x_{2},x_{3})$ can take any 8 values (000, 100, 010, 001, 110, 101, 011 or 111) from $t=3$;
  \item $(x_{1},x_{2},x_{3},x_{4})$ can take any 16 values from $t=4$.
\end{itemize}
Suppose $\xvec^0=[0,0,0,0]$ and $\xvec^T=[1,1,1,1]$. Let $u_{3}(0)=1,u_{3}(1)=1,u_{3}(2)=0,u_{3}(3)=0$, we have $\xvec(1)=[0,0,1,0],\xvec(2)=[1,0,0,0],\xvec(3)=[1,1,0,0]$ and $\xvec(4)=[1,1,1,1]=\xvec^T$.
Next, suppose $\xvec^0=[0,0,0,0]$ and $\xvec^T=[1,1,0,1]$. Let $u_{3}(0)=1,u_{3}(1)=1,u_{3}(2)=0,u_{3}(3)=1$, we have $\xvec(1)=[0,0,1,0],\xvec(2)=[1,0,0,0],\xvec(3)=[1,1,0,0]$ and $\xvec(4)=[1,1,0,1]=\xvec^T$. Here, $U=\{x_{3}\}$ is the minimum control node set.

On the other hand, let $U=\{x_{1}\}$, and then we find that (Fig. \ref{fig:xor2b})
\begin{itemize}
  \item $x_{1}$ can take any 2 values (0 or 1) from $t=1$;
  \item $(x_{1},x_{2})$ can take any 4 values (00, 10, 01 or 11) from $t=2$;
  \item $(x_{1},x_{2},x_{3},x_{4})$ can take 8 values from $t=3$, where the value of $(x_{3},x_{4})$ depends on $\xvec^0$ and $(x_{3}(t),x_{4}(t))$ can take 2 values $(x_{3}(t-1),x_{4}(t-1))$ or $(\overline{x_{3}(t-1)},\overline{x_{4}(t-1)})$ from $t=3$.
\end{itemize}
Suppose $\xvec^0=[0,0,0,0]$ and $\xvec^T=[1,1,0,1]$. If $U=\{x_{1}\}$, this BN cannot reach $\xvec^T=[1,1,0,1]$ from the initial state $\xvec^0=[0,0,0,0]$.

\begin{table}[htp]
\caption{All possible states from $t=0$ to $t=4$ in Example \ref{ex:xor3}, where $U=\{x_{3}\}$.}\label{table1}
\begin{center}
\begin{tabular}{|c|cccc|}
\hline
&$x_{1}$&$x_{2}$&$x_{3}$&$x_{4}$\\
\hline
$t=0$&$x_{1}(0)$&$x_{2}(0)$&$x_{3}(0)$&$x_{4}(0)$\\
\hline
$t=1$&$x_{1}(0)\oplus x_{3}(0)$&$x_{1}(0)\oplus x_{4}(0)$&$0$&$x_{2}(0)\oplus x_{4}(0)$\\
$t=1$&{\color{blue}$x_{1}(0)\oplus x_{3}(0)$}&{\color{blue}$x_{1}(0)\oplus x_{4}(0)$}&{\color{blue}$1$}&{\color{blue}$x_{2}(0)\oplus x_{4}(0)$}\\
\hline
$t=2$&$x_{1}(0)\oplus x_{3}(0)$&$x_{1}(0)\oplus x_{2}(0)\oplus x_{3}(0)\oplus x_{4}(0)$&$0$&$x_{1}(0)\oplus x_{2}(0)$\\
$t=2$&$x_{1}(0)\oplus x_{3}(0)$&$x_{1}(0)\oplus x_{2}(0)\oplus x_{3}(0)\oplus x_{4}(0)$&$1$&$x_{1}(0)\oplus x_{2}(0)$\\
$t=2$&{\color{blue}$\overline{x_{1}(0)\oplus x_{3}(0)}$}&{\color{blue}$x_{1}(0)\oplus x_{2}(0)\oplus x_{3}(0)\oplus x_{4}(0)$}&{\color{blue}$0$}&{\color{blue}$x_{1}(0)\oplus x_{2}(0)$}\\
$t=2$&$\overline{x_{1}(0)\oplus x_{3}(0)}$&$x_{1}(0)\oplus x_{2}(0)\oplus x_{3}(0)\oplus x_{4}(0)$&$1$&$x_{1}(0)\oplus x_{2}(0)$\\
\hline
$t=3$&$x_{1}(0)\oplus x_{3}(0)$&$x_{2}(0)\oplus x_{3}(0)$&$0$&$x_{3}(0)\oplus x_{4}(0)$\\
$t=3$&$x_{1}(0)\oplus x_{3}(0)$&$x_{2}(0)\oplus x_{3}(0)$&$1$&$x_{3}(0)\oplus x_{4}(0)$\\
$t=3$&$\overline{x_{1}(0)\oplus x_{3}(0)}$&$x_{2}(0)\oplus x_{3}(0)$&$0$&$x_{3}(0)\oplus x_{4}(0)$\\
$t=3$&$\overline{x_{1}(0)\oplus x_{3}(0)}$&$x_{2}(0)\oplus x_{3}(0)$&$1$&$x_{3}(0)\oplus x_{4}(0)$\\
$t=3$&{\color{blue}$\overline{x_{1}(0)\oplus x_{3}(0)}$}&{\color{blue}$\overline{x_{2}(0)\oplus x_{3}(0)}$}&{\color{blue}$0$}&{\color{blue}$x_{3}(0)\oplus x_{4}(0)$}\\
$t=3$&$\overline{x_{1}(0)\oplus x_{3}(0)}$&$\overline{x_{2}(0)\oplus x_{3}(0)}$&$1$&$x_{3}(0)\oplus x_{4}(0)$\\
$t=3$&$x_{1}(0)\oplus x_{3}(0)$&$\overline{x_{2}(0)\oplus x_{3}(0)}$&$0$&$x_{3}(0)\oplus x_{4}(0)$\\
$t=3$&$x_{1}(0)\oplus x_{3}(0)$&$\overline{x_{2}(0)\oplus x_{3}(0)}$&$1$&$x_{3}(0)\oplus x_{4}(0)$\\
\hline
$t=4$&$x_{1}(0)\oplus x_{3}(0)$&$x_{1}(0)\oplus x_{4}(0)$&$0$&$x_{2}(0)\oplus x_{4}(0)$\\
$t=4$&$x_{1}(0)\oplus x_{3}(0)$&$x_{1}(0)\oplus x_{4}(0)$&$1$&$x_{2}(0)\oplus x_{4}(0)$\\
$t=4$&$\overline{x_{1}(0)\oplus x_{3}(0)}$&$x_{1}(0)\oplus x_{4}(0)$&$0$&$x_{2}(0)\oplus x_{4}(0)$\\
$t=4$&$\overline{x_{1}(0)\oplus x_{3}(0)}$&$x_{1}(0)\oplus x_{4}(0)$&$1$&$x_{2}(0)\oplus x_{4}(0)$\\
$t=4$&$\overline{x_{1}(0)\oplus x_{3}(0)}$&$\overline{x_{1}(0)\oplus x_{4}(0)}$&$0$&$x_{2}(0)\oplus x_{4}(0)$\\
$t=4$&$\overline{x_{1}(0)\oplus x_{3}(0)}$&$\overline{x_{1}(0)\oplus x_{4}(0)}$&$1$&$x_{2}(0)\oplus x_{4}(0)$\\
$t=4$&$x_{1}(0)\oplus x_{3}(0)$&$\overline{x_{1}(0)\oplus x_{4}(0)}$&$0$&$x_{2}(0)\oplus x_{4}(0)$\\
$t=4$&$x_{1}(0)\oplus x_{3}(0)$&$\overline{x_{1}(0)\oplus x_{4}(0)}$&$1$&$x_{2}(0)\oplus x_{4}(0)$\\
$t=4$&{\color{blue}$\overline{x_{1}(0)\oplus x_{3}(0)}$}&{\color{blue}$\overline{x_{1}(0)\oplus x_{4}(0)}$}&{\color{blue}$0$}&{\color{blue}$\overline{x_{2}(0)\oplus x_{4}(0)}$}\\
$t=4$&{\color{blue}$\overline{x_{1}(0)\oplus x_{3}(0)}$}&{\color{blue}$\overline{x_{1}(0)\oplus x_{4}(0)}$}&{\color{blue}$1$}&{\color{blue}$\overline{x_{2}(0)\oplus x_{4}(0)}$}\\
$t=4$&$x_{1}(0)\oplus x_{3}(0)$&$\overline{x_{1}(0)\oplus x_{4}(0)}$&$0$&$\overline{x_{2}(0)\oplus x_{4}(0)}$\\
$t=4$&$x_{1}(0)\oplus x_{3}(0)$&$\overline{x_{1}(0)\oplus x_{4}(0)}$&$1$&$\overline{x_{2}(0)\oplus x_{4}(0)}$\\
$t=4$&$x_{1}(0)\oplus x_{3}(0)$&$x_{1}(0)\oplus x_{4}(0)$&$0$&$\overline{x_{2}(0)\oplus x_{4}(0)}$\\
$t=4$&$x_{1}(0)\oplus x_{3}(0)$&$x_{1}(0)\oplus x_{4}(0)$&$1$&$\overline{x_{2}(0)\oplus x_{4}(0)}$\\
$t=4$&$\overline{x_{1}(0)\oplus x_{3}(0)}$&$x_{1}(0)\oplus x_{4}(0)$&$0$&$\overline{x_{2}(0)\oplus x_{4}(0)}$\\
$t=4$&$\overline{x_{1}(0)\oplus x_{3}(0)}$&$x_{1}(0)\oplus x_{4}(0)$&$1$&$\overline{x_{2}(0)\oplus x_{4}(0)}$\\
\hline
\end{tabular}
\end{center}
\end{table}
\end{example}

For Example \ref{ex:xor3}, according to Theorem \ref{thm:xor-upper}, we have $W=\{x_{3}\}$, $U=\{x_{1},x_{2},x_{3}\}$, $V_{1}^{0}=\{x_{2}\}$, $V_{1}^{1}=\emptyset$, $V_{1}^{2}=\{x_{1}\}$, $V_{2}=\{x_{3},x_{4}\}$, $V_{3}=\emptyset$ and $\phi(2)=4$. Similarly, suppose $\xvec^0=[0,0,0,0]$ and $\xvec^T=[1,1,1,1]$. Then, we have $\xvec^2=[0,0,0,0]$ (at $t=2$). At $t=0$, we let $u_2(0)=1$ and let $u_i(0)=0$ for the other control nodes.
At $t=1$, we let $u_1(1)=u_2(1)=1$ and $u_3(1)=0$.
Then, we have the following.
\begin{center}
\begin{tabular}{l|llll}
\hline
$\xvec^0$ ($t=0$) & 0 & 0 & 0 & 0 \\
$\xvec^1$ ($t=1$) & 0 & 0 & 0 & 0 \\
$\xvec^2$ ($t=2$) & 0 & 0 & 0 & 0 \\
\hline
$\xvec^3$ ($t=1$) & 0 & 1 & 0 & 0 \\
$\xvec^4$ ($t=2$) & 0 & 0 & 1 & 1 \\
$\xvec^5$ ($t=2$) & 1 & 1 & 1 & 1 \\
\hline
\end{tabular}
\end{center}
where $\xvec^1,\ldots,\xvec^5$ are defined as in Example \ref{ex:xor1}, except that the control signals at $t=0$ and $t=1$ differ.

Therefore, we consider reducing the size of the control node set by increasing the time to reach the target state $x(t)=\xvec^T$ to some extent.

\begin{theorem}\label{thm:xor-time}
For any 2-2-XOR-BN, the set of control nodes is not less than $\frac{n}{t^{*}}$ and not greater than $n-t^{*}+1$, where $t^{*}$ is the minimum time that any initial state $\xvec^0$ can reach any target state $\xvec^T$ and $t^{*}\in[1,n]$.
\end{theorem}
(Proof)
If a 2-2-XOR-BN is controllable, assuming $t^{*}$ is the minimum time for any initial state $\xvec^0$ to reach any target state $\xvec^T$, then the following sets are not empty,
\begin{eqnarray*}
U_{1} & = & U,\\
U_{2} & = & \{x_{j_{2}}|x_{j_{2}}\in\Gamma^{+}(x_{j_{1}})\setminus U_{1} ~\land~ x_{j_{1}}\in U_{1}\},\\
U_{3} & = & \{x_{j_{3}}|x_{j_{3}}\in\Gamma^{+}(x_{j_{2}})\setminus(U_{1}\cup U_{2}) ~\land~ x_{j_{2}}\in U_{2}\},\\
&\vdots&\\
U_{t^{*}} & = & \{x_{j_{t^{*}}}|x_{j_{t^{*}}}\in\Gamma^{+}(x_{j_{t^{*}-1}})\setminus(\cup_{i=1}^{t^{*}-1} U_{i}) ~\land~ x_{j_{t^{*}-1}}\in U_{t^{*}-1}\},
\end{eqnarray*}
$\cup_{i=1}^{t^{*}}U_{i}=V$. Due to the property of exclusive OR, that is, $x(t+1)$ can take any value 0 or 1 as long as there is an incoming node $x_{i}$ ($x_{i}\in\Gamma^{-}(x)$) that can take any value 0 or 1 at time $t$, we can see that by setting $U_{1}$ as the control node set, the nodes in $U_{i},~i=2,3,\ldots,t^{*}$ can take any two values 0 or 1 from $t=i$, which means that it is possible to make the 2-2-XOR-BN controllable by controlling the nodes in $U_{1}$.

Hence, for each $U_{i},~i=2,\ldots,t^{*}$, we have $|U_{i}|>1$ and then
\begin{eqnarray*}
n & = & |U_1| + |U_2| + \cdots+ |U_{t^{*}}|\\
& \geq & |U_1| + 1 + \cdots+ 1\\
& = & |U_1|+t^{*}-1,
\end{eqnarray*}
which means that $|U_{1}|\leq n-t^{*}+1$.

On the other hand, we assume without loss of generality $x_{1},\ldots,x_{|U_{1}|}$ are control nodes.  Then we note that for any initial state $\xvec^0$,
\begin{itemize}
  \item at time $t=1$, there are $2^{|U_{1}|}$ different states, because there are $|U_{1}|$ control nodes;
  \item at time $t=2$, there are at most $2^{2|U_{1}|}$ different states, because the number of different values for $(x_{|U_{1}|+1}(2),\ldots,x_{n}(2))$ is at most $2^{|U_{1}|}$;
  \item at time $t=3$, there are at most $2^{3|U_{1}|}$ different states, because the number of different values for $(x_{|U_{1}|+1}(3),\ldots,x_{n}(3))$ is at most $2^{2|U_{1}|}$;
  \item $\ldots$
  \item at time $t=t^{*}$, there are at most $2^{t^{*}|U_{1}|}$ different states, because the number of different values for $(x_{|U_{1}|+1}(t^{*}),\ldots,x_{n}(t^{*}))$ is at most $2^{(t^{*}-1)|U_{1}|}$.
\end{itemize}
Because we assume that $t^{*}$ is the minimum time for any initial state $\xvec^0$ to reach any target state $\xvec^T$, then $2^{t^{*}|U_{1}|}\geq2^{n}$, i.e., $|U_{1}|\geq\frac{n}{t^{*}}$.
\qed

\bigskip

\begin{remark}\label{remark2}
The above result also applies to $k$-$k$-XOR-BNs ($k\geq2$). That is, for any $k$-$k$-XOR-BN, the size of the control node set is not less than $\frac{n}{t^{*}}$ and not greater than $n-t^{*}+1$, where $t^{*}$ is the minimum time that any initial state $\xvec^0$ can reach any target state $\xvec^T$ and $t^{*}\in[1,n]$.
\end{remark}

\begin{remark}
According to Theorem \ref{thm:xor-time}, if $t^{*}=1$, then the size of the control node set is $n$. On the other hand, it does not mean that $t^{*}$ can take any value from $[3,n]$. To illustrate why $t^{*}$ cannot take any value from $[3,n]$, we consider the 2-2-XOR-BN in Example \ref{ex:xor1}, where $n=6$ and
\begin{eqnarray*}
x_{1}(t+1) & = & x_{1}(t) \lxor x_{3}(t),\\
x_{2}(t+1) & = & x_{1}(t) \lxor x_{3}(t),\\
x_{3}(t+1) & = & x_{2}(t) \lxor x_{4}(t),\\
x_{4}(t+1) & = & x_{2}(t) \lxor x_{5}(t),\\
x_{5}(t+1) & = & x_{4}(t) \lxor x_{6}(t),\\
x_{6}(t+1) & = & x_{5}(t) \lxor x_{6}(t).
\end{eqnarray*}
If $t^{*}$ could take the value 6, then $|U_{1}|=1$ according to Theorem \ref{thm:xor-time}, and there are six cases as follows
(each case can be verified by exhaustive examination of states of $x_i(t)$
with $t=0,\ldots,6$ for $x_i \in U_1$):
\begin{itemize}
  \item When $U_{1}=\{x_{1}\}$, it can be seen that we cannot drive this BN from any initial state $\xvec^0$ to any target state $\xvec^T$ at $t=6$. For example, let $\xvec^0=[0,0,0,0,0,0]$ and $\xvec^T=[0,0,0,1,0,0]$.
  \item When $U_{1}=\{x_{2}\}$, it can be seen that we cannot drive this BN from any initial state $\xvec^0$ to any target state $\xvec^T$ at $t=6$. For example, let $\xvec^0=[0,0,0,0,0,0]$ and $\xvec^T=[0,0,0,1,0,0]$.
  \item When $U_{1}=\{x_{3}\}$, it can be seen that we cannot drive this BN from any initial state $\xvec^0$ to any target state $\xvec^T$ at $t=6$. For example, let $\xvec^0=[0,0,0,0,0,0]$ and $\xvec^T=[1,0,0,0,0,0]$.
  \item When $U_{1}=\{x_{4}\}$, it can be seen that we cannot drive this BN from any initial state $\xvec^0$ to any target state $\xvec^T$ at $t=6$. For example, let $\xvec^0=[0,0,0,0,0,0]$ and $\xvec^T=[0,0,1,0,0,0]$.
  \item When $U_{1}=\{x_{5}\}$, it can be seen that we cannot drive this BN from any initial state $\xvec^0$ to any target state $\xvec^T$ at $t=6$. For example, let $\xvec^0=[0,0,0,0,0,0]$ and $\xvec^T=[0,0,0,1,0,0]$.
  \item When $U_{1}=\{x_{6}\}$, it can be seen that we cannot drive this BN from any initial state $\xvec^0$ to any target state $\xvec^T$ at $t=6$. For example, let $\xvec^0=[0,0,0,0,0,0]$ and $\xvec^T=[1,0,0,0,0,0]$.
\end{itemize}
Therefore, $t^{*}$ cannot take the value 6.
\end{remark}

Next, we consider $k$-$k$-XOR-BNs when $k$ is an odd number with $k \geq 3$.
For example, consider the case of $k=3$ and $n=4m$ ($d=3$ and $n=4$ in \cite{guo22}).
In this case, we consider the following BN:
\begin{eqnarray*}
x_{4h+5}(t+1) & = & x_{4h+2}(t) \lxor x_{4h+3}(t) \lxor x_{4h+4}(t),\\
x_{4h+6}(t+1) & = & x_{4h+1}(t) \lxor x_{4h+3}(t) \lxor x_{4h+4}(t),\\
x_{4h+7}(t+1) & = & x_{4h+1}(t) \lxor x_{4h+2}(t) \lxor x_{4h+4}(t),\\
x_{4h+8}(t+1) & = & x_{4h+1}(t) \lxor x_{4h+2}(t) \lxor x_{4h+3}(t),
\end{eqnarray*}
where the subscript is given with modulo $n=4m$ ($+1$).
Clearly, this BN is a 3-3-XOR-BN.
From Lemma 14 of \cite{guo22},
the entropy is preserved from
$[x_{4h+1}(t),x_{4h+2}(t),x_{4h+3}(t),x_{4h+4}(t)]$
to
$[x_{4h+5}(t+1),x_{4h+6}(t+1),x_{4h+7}(t+1),x_{4h+8}(t+1)]$.
We let $U=\{x_1,x_2,x_3,x_4\}$.
Then,
we can drive the BN to any desired state at time step $m=\frac{n}{4}$
by giving appropriate signal to $[u_1(t),u_2(t),u_3(t),u_4(t)]$.
By generalizing this construction, we have:

\begin{theorem}\label{thm:4-4-xor}
For each odd $k \geq 3$,
there exists a $k$-$k$-XOR-BN with $n=(k+1)m$ nodes
with a control node set of size $k+1$.
\end{theorem}

This theorem can be modified for some BNs with threshold functions.

Then we extend Theorem \ref{thm:xor-upper} to $k$-$k$-XOR-BNs, $k\geq2$.
\begin{theorem}\label{theorem4}
Suppose that $n=[k(k-1)+1]m$, where $k\geq2$ and $(m\mod k)=0$. Then, for any $k$-$k$-XOR-BN,
there exists a control node set of size $\left[1- \frac{k-1}{k(k^{2}-k+1)}\right]n$.
\end{theorem}
(Proof)
Similar to Lemma \ref{lemma1}, we first use the greedy procedure to find a required $V_{1}$, but the difference is that there are at most $k(k-1)$ such $x_{j}$s at STEP 4 because of the constraints on indegree and outdegree, so the size of $S$ decreases by at most $k(k-1)+1$ for each iteration. Hence, we can get a required  $V_{1}\subseteq V$ with $|V_{1}|=\frac{1}{k(k-1)+1}n=m$ such that $\Gamma^{+}(x_i) \cap \Gamma^+(x_j) = \emptyset$ holds for any $x_i \neq x_j$ in $V_1$.
Then, we also partition $V$ into three disjoint sets $V_{1},V_{2}$, and $V_{3}$, where
$V_2 = \{x_j \mid (x_i,x_j) \in E ~\land~ x_i \in V_1 ~\land~ x_j \notin V_1 \}$ and
$V_3 = V \setminus (V_1 \cup V_2)$. Here, $|V_{1}|=m$, $|V_2|\geq (k-1)m$ and $|V_3| \leq[k(k-2)+1]m$.

Let $N(V,F)$ be a given $k$-$k$-XOR-BN.
We assume without loss of generality $V_1 = \{x_1,\ldots,x_m\}$
(by appropriately permuting indices of $x_i$s).

We partition $V_1$ into $k+1$ disjoint sets
$V_1^0,V_1^1,\ldots,V_1^k$, where
\begin{eqnarray*}
V_1^0 & = & \{x_i \in V_1 \mid |\Gamma^+(x_i) \cap V_1|=0\},\\
V_1^1 & = & \{x_i \in V_1 \mid |\Gamma^+(x_i) \cap V_1|=1\},\\
&\cdots&\\
V_1^{k-1} & = & \{x_i \in V_1 \mid |\Gamma^+(x_i) \cap V_1|=k-1\},\\
V_1^{k} & = & \{x_i \in V_1 \mid |\Gamma^+(x_i) \cap V_1|=k\}.
\end{eqnarray*}
For each node $x_i \in V_1^0 \cup V_1^1\cup\cdots\cup V_1^{k-1}$,
we let $\phi(i)=j_k$ where $\Gamma^+(x_i)=\{x_{j_1},x_{j_2},\ldots,x_{j_k}\}$ and $j_1 < j_2<\cdots< j_{k}$.
Clearly, $x_{\phi(i)} \in V_2$ holds.
$\phi(i)$ is introduced so that each $x_{\phi(i)}$ takes the desired value
at $t=2$ by giving an appropriate control to $x_i$ at $t=1$.

Then we define $W^{0},W^{1},\ldots,W^{k-2}$ by
\begin{eqnarray*}
W^0 & = & \{x_{j_1},x_{j_2},\ldots,x_{j_{k-1}} \mid \Gamma^+(x_i)=\{x_{j_1},x_{j_2},\ldots,x_{j_k}\}\wedge x_{i}\in V_{1}^{0}\wedge j_1 < j_2<\cdots< j_{k}\},\\
W^1 & = & \{x_{j_2},\ldots,x_{j_{k-1}} \mid \Gamma^+(x_i)=\{x_{j_1},x_{j_2},\ldots,x_{j_k}\}\wedge x_{i}\in V_{1}^{1}\wedge j_1 < j_2<\cdots< j_{k}\},\\
&\cdots&\\
W^{k-2} & = & \{x_{j_{k-1}} \mid \Gamma^+(x_i)=\{x_{j_1},x_{j_2},\ldots,x_{j_k}\}\wedge x_{i}\in V_{1}^{k-2}\wedge j_1 < j_2<\cdots< j_{k}\}.
\end{eqnarray*}
Let $W=W^{0}\cup W^{1}\cup\cdots\cup W^{k-2}$. It is seen that $W\subseteq V_{2}, |W|=(k-1)|V_{1}^{0}|+(k-2)|V_{1}^{1}|+\cdots+|V_{1}^{k-2}|$ and $|V_{2}|=k|V_{1}^{0}|+(k-1)|V_{1}^{1}|+\cdots+2|V_{1}^{k-2}|+|V_{1}^{k-1}|$. Hence, $|W| \leq {\frac {k-1} k} |V_2|$.
Note also that $x_{\phi(i)} \notin W$ for any $x_i \in V_1^0 \cup V_1^1\cup\cdots\cup V_1^{k-1}$.

Then, we let $U = V_1 \cup W \cup V_3$ be the set of control nodes.
From $|V_1|=m$, $|V_2| \geq (k-1)m$, and $|V_3|=n-|V_1|-|V_2|$,
we have
\begin{eqnarray*}
|U| & = & |V_1| + |W| + |V_3|\\
& = & |V_1| + |W| + n - |V_1| - |V_2|\\
& = & n - |V_2| + |W|\\
& \leq & n - {\frac 1 k}|V_2| \\
& \leq & n - {\frac {k-1} k}m \\
& = &\left[1- \frac{k-1}{k(k^{2}-k+1)}\right]n.
\end{eqnarray*}

Next, we show that $U$ satisfies the conditions of control node set.
Let $\xvec^0$ and $\xvec^T$ be specified initial and target state vectors,
respectively.
Let $\xvec^1 = F(\xvec^0)$ and $\xvec^2=F(\xvec^1)$ (without any control).
Then, controls are given as below.

At $t=0$,
for each node $x_i \in V_1^0 \cup V_1^1\cup\cdots\cup V_1^{k-1}$ with
$(\xvec^2)_{\phi(i)} \neq (\xvec^T)_{\phi(i)}$,
we let $u_i(0) = 1$.
For any other control node $x_i$,
we let $u_i(0) = 0$.
Let the resulting state vectors at $t=1$ and $t=2$
be $\xvec^3$ and $\xvec^4$, respectively.
Then, $(\xvec^4)_{\phi(i)} = (\xvec^T)_{\phi(i)}$ holds for
all $\phi(i)$.

At $t=1$,
for each node $x_i \in U$,
we let $u_i(1)=1$ if $(\xvec^4)_i \neq (\xvec^T)_i$,
otherwise we let $u_i(1)=0$.
Then, the resulting state vector $\xvec^5$ (at $t=2$) satisfies
$\xvec^5 = \xvec^T$.
\qed

\bigskip
\begin{remark}
Note that Theorem \ref{thm:xor-upper} is a special case of Theorem \ref{theorem4}, that is,
by letting $k=2$, the general upper bound on the size of the control set of Theorem \ref{theorem4} becomes $\frac{5}{6}n$.
\end{remark}
\section{Boolean networks consisting of simple AND/OR functions}
Suppose that the indegree and outdegree of each node are both 2,
and all functions of $F$ are simple AND functions
(i.e., with the form of $x \land y$ (no negations)).
Such a BN is referred to as a simple 2-2-AND-BN.

\begin{proposition}\label{proposition4}
For any $n=2m$ with a positive integer $m$,
there exists a simple 2-2-AND-BN for which the size of the minimum control node set is $n$.
\end{proposition}
(Proof)
We construct a simple 2-2-AND-BN by
\begin{eqnarray*}
x_{2i-1}(t+1) & = & x_{2i-1}(t) \land x_{2i}(t),\\
x_{2i}(t+1) & = & x_{2i-1}(t) \land x_{2i}(t).
\end{eqnarray*}
Suppose that $\xvec^0=[0,0,\ldots,0]$ and
$\xvec^T=[1,1,\ldots,1]$.
It is straightforward to see that the size of the minimum control node set must be $n$
(i.e., all nodes are control nodes).
\qed

In the following,
we show best case upper bounds for 2-AND-BN and 2-2-AND-BN,
where the latter will be generalized for $k$-$k$-AND-BN.
The basic idea is to use a kind of shift-register,
by which we can propagate signals given to the control node
(e.g., $x_1$) to the other nodes, where
some additional nodes are required to satisfy
degree constraints.

\begin{figure}[th]
\begin{center}
\includegraphics[width=10cm]{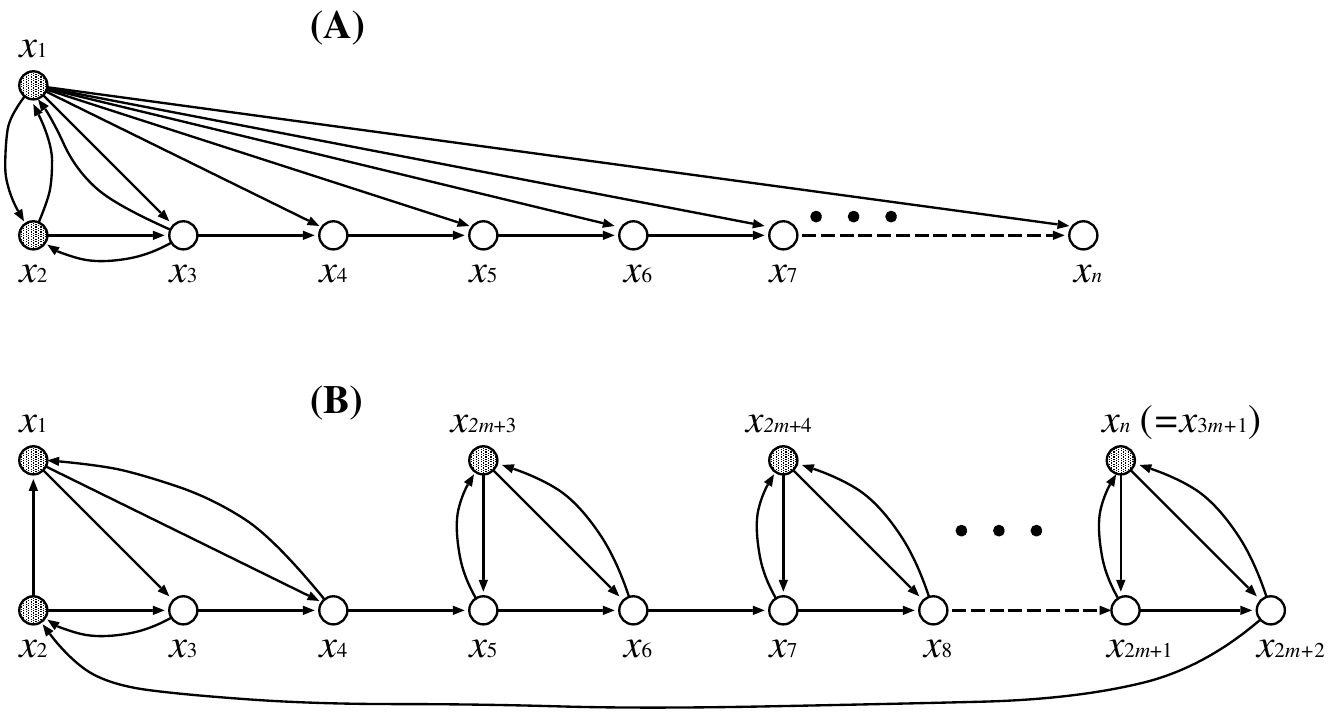}
\caption{BNs constructed in the proofs of Proposition \ref{prop:2-and-upper}
and Proposition \ref{prop:2-2-and-upper}.}
\label{fig:2-2-and-upper}
\end{center}
\end{figure}

\begin{proposition}
For any integer $n > 2$,
there exists a 2-AND-BN (i.e., every node has indegree 2)
for which the size of the minimum control node set is $2$.
\label{prop:2-and-upper}
\end{proposition}
(Proof)
We construct a 2-AND-BN by
\begin{eqnarray*}
x_1(t+1) & = & x_2(t) \land x_3(t),\\
x_2(t+1) & = & x_1(t) \land x_3(t),\\
x_3(t+1) & = & x_1(t) \land x_2(t),\\
x_4(t+1) & = & x_1(t) \land x_3(t),\\
& \vdots & \\
x_n(t+1) & = & x_1(t) \land x_{n-1}(t)
\end{eqnarray*}
with letting $x_1$ and $x_2$ be control nodes
(see Fig.~\ref{fig:2-2-and-upper}(A)).
By using control signals,
we let $x_1(t)=1$ for all $t=1,\ldots,n-2$,
$x_1(n-1)=x^T_1$,
and $x_2(t) = x^T_{n-t+1}$ for $t=1,\ldots,n-1$.
Then, it is straightforward to see that $\xvec(n-1)=\xvec^T$.
\qed

\bigskip

\begin{proposition}\label{proposition6}
For any integer $n = 3m+1$ with a positive integer $m>1$,
there exists a simple 2-2-AND-BN
for which the size of the minimum control node set is $m+1 = {\frac {n-1} 3} + 1$.
\label{prop:2-2-and-upper}
\end{proposition}
(Proof)
We construct a simple 2-2-AND-BN by
\begin{eqnarray*}
x_1(t+1) & = & x_2(t) \land x_4(t),\\
x_2(t+1) & = & x_3(t) \land x_{2m+2}(t),\\
x_3(t+1) & = & x_1(t) \land x_2(t),\\
x_4(t+1) & = & x_1(t) \land x_3(t),\\
x_5(t+1) & = & x_{2m+3}(t) \land x_4(t),\\
x_6(t+1) & = & x_{2m+3}(t) \land x_5(t),\\
x_7(t+1) & = & x_{2m+4}(t) \land x_6(t),\\
x_8(t+1) & = & x_{2m+4}(t) \land x_7(t),\\
& \vdots & \\
x_{n}(t+1) & = & x_{2m+1}(t) \land x_{2m+2}(t),\\
x_{2m+1}(t+1) & = & x_{n}(t) \land x_{2m}(t),\\
x_{2m+2}(t+1) & = & x_{n}(t) \land x_{2m+1}(t)
\end{eqnarray*}
with letting $x_1$, $x_2$, and $x_{2m+2+i}$ ($i=1,\dots,m-1$)
be control nodes
(see Fig.~\ref{fig:2-2-and-upper}(B)).
By using control signals,
we let $x_1(t)=1$ and $x_{2m+2+i}(t)=1$ ($i=1,\ldots,m-1$)
for all $t=1,\ldots,2m$,
$x_1(2m+1)=x^T_1$.
$x_{2m+2+i}(2m+1)=x_{2m+2+i}^T$ ($i=1,\ldots,m-1$),
and $x_2(t) = x^T_{2m+3-t}$ for $t=1,\ldots,2m+1$.
Then, it is straightforward to see that $\xvec(2m+1)=\xvec^T$.
\qed

\bigskip

In the following, we show a general lower bound on the size of the control node set for the case of simple
2-2-AND-BNs.

\begin{theorem}\label{thm:2-2-and}
For any simple 2-2-AND-BN, the size of the control node set is at least $\frac{n}{3}$.
\end{theorem}
(Proof)
Note that for any node $x_{i}$ in a simple 2-2-AND-BN, we have $|\Gamma^{+}(x_{i})|=|\Gamma^{-}(x_{i})|=2$.

Then for any $x_{i}\notin U$, there are five cases:
(i) $|\Gamma^{+}(x_{i})\cap U|=2$; (ii) $|\Gamma^{+}(x_{i})\cap U|=1$ and $\Gamma^{+}(x_{i})\cap\{x_{i}\}=\emptyset$;
(iii) $|\Gamma^{+}(x_{i})\cap U|=1$ and $x_{i}\in\Gamma^{+}(x_{i})$;
(iv) $\Gamma^{+}(x_{i})\cap U=\emptyset$ and $\Gamma^{+}(x_{i})\cap\{x_{i}\}=\emptyset$;
(v) $\Gamma^{+}(x_{i})\cap U=\emptyset$ and $x_{i}\in\Gamma^{+}(x_{i})$.
First, it is seen that if a simple 2-2-AND-BN is controllable, then any node $x_{i}\notin U$ will not encounter Cases (iii) and (v), that is, $x_{i}\notin\Gamma^{+}(x_{i})$. Suppose there is a node $x_{i}\notin U$, where $x_{i}\in\Gamma^{+}(x_{i})$. If $x_{i}(0)=0$, then $x_{i}(t)$ will always be 0, which contradicts the fact that the simple 2-2-AND-BN is controllable.

Now, we divide any node $x_{i}\notin U$ in a controllable simple 2-2-AND-BN into three disjoint sets  $V_0$, $V_1$, and $V_2$, where
\begin{eqnarray*}
V_{0} & = & \{x_{i}\mid x_{i}\notin U ~\land~ \Gamma^{+}(x_{i})\cap U=\emptyset~\land~ \Gamma^{+}(x_{i})\cap\{x_{i}\}=\emptyset\},\\
V_{1} & = & \{x_{i}\mid x_{i}\notin U ~\land~ |\Gamma^{+}(x_{i})\cap U|=1 ~\land~ \Gamma^{+}(x_{i})\cap\{x_{i}\}=\emptyset\},\\
V_{2} & = & \{x_{i}\mid x_{i}\notin U ~\land~|\Gamma^{+}(x_{i})\cap U|=2\}.
\end{eqnarray*}
Similarly, we divide any node $x_{j}\in U$ in a controllable simple 2-2-AND-BN into three disjoint sets  $W_0$, $W_1$, and $W_2$, where
\begin{eqnarray*}
W_{0} & = & \{x_{j}\mid x_{j}\in U ~\land~ \Gamma^{+}(x_{j})\cap U=\emptyset\},\\
W_{1} & = & \{x_{j}\mid x_{j}\in U ~\land~ |\Gamma^{+}(x_{j})\cap U|=1\},\\
W_{2} & = & \{x_{j}\mid x_{j}\in U ~\land~ |\Gamma^{+}(x_{j})\cap U|=2\}.
\end{eqnarray*}
Notably, $|V|=|V_{0}|+|V_{1}|+|V_{2}|+|U|=|V_{0}|+|V_{1}|+|V_{2}|+|W_{0}|+|W_{1}|+|W_{2}|$.

Then, it is seen that if a simple 2-2-AND-BN is controllable, then for any node $x_{i}\notin U$, we have $\Gamma^{-}(x_{i})\cap(V_{1}\cup W_{1})\neq\emptyset$. Suppose there is a node $x_{i}\notin U$, where $\Gamma^{-}(x_{i})\cap V_{1}=\emptyset$ and $\Gamma^{-}(x_{i})\cap W_{1}=\emptyset$. We assume without loss of generality $x_{i}(t+1)=x_{i_{1}}(t)\wedge x_{i_{2}}(t)$, where $x_{i_{1}},x_{i_{2}}\notin V_{1}\cup W_{1}$. Clearly, $x_{i_{1}},x_{i_{2}}\notin V_{2}\cup W_{2}$, because for any $x_{l}\in V_{2}\cup W_{2}$,  it is definitely not an incoming node for any $x_{i}\notin U$. Then we assume without loss of generality $\Gamma^{+}(x_{i_{1}})=\{x_{i},x_{k_{1}}\}$, $\Gamma^{+}(x_{i_{2}})=\{x_{i},x_{k_{2}}\}$, where $x_{k_{1}},x_{k_{2}}\notin U$. Here, $x_{k_{1}},x_{k_{2}}$ can be the same, that is, the situation where $\Gamma^{+}(x_{i_{1}})=\Gamma^{+}(x_{i_{2}})$ is possible, and $x_{i_{1}}$ (resp., $x_{i_{2}}$) can belong to $U$ or not belong to $U$. Then for any initial state $\xvec^0$, the target state of the form $x_{i}=0,x_{k_{1}}=x_{k_{2}}=1$ cannot be reached, which contradicts the fact that the simple 2-2-AND-BN is controllable.

Therefore, if a simple 2-2-AND-BN is controllable, then for any $x_{i}\notin U$, there are three cases:
\begin{description}
  \item[(i)] $|\Gamma^{+}(x_{i})\cap U|=2$ and $\Gamma^{-}(x_{i})\cap(V_{1}\cup W_{1})\neq\emptyset$;
  \item[(ii)] $|\Gamma^{+}(x_{i})\cap U|=1$, $\Gamma^{+}(x_{i})\cap\{x_{i}\}=\emptyset$ and $\Gamma^{-}(x_{i})\cap(V_{1}\cup W_{1})\neq\emptyset$;
  \item[(iii)] $\Gamma^{+}(x_{i})\cap U=\emptyset$, $\Gamma^{+}(x_{i})\cap\{x_{i}\}=\emptyset$ and $\Gamma^{-}(x_{i})\cap(V_{1}\cup W_{1})\neq\emptyset$.
\end{description}

Based on the above analysis, for any node $x_{l}\in V_{1}\cup W_{1}$ (resp., $x_{l}\in V_{2}\cup W_{2}$), there is an outgoing node that belongs to $U$ (resp., there are two outgoing nodes that belong to $U$), that is, there is an edge pointing from $x_{l}$ to the control node (resp., there are two edges pointing from $x_{l}$ to the control nodes). It is known that the total number of in-degrees of control nodes is $2|U|$, and for any node $x_{l}\in V_{0}\cup W_{0}$, there is no outgoing node belonging to $U$. Hence, we can get Eq. (\ref{e1}).

On the other hand, if a simple 2-2-AND-BN is controllable, then for any node $x_{i}\in V_{0}\cup V_{1}\cup V_{2}$, there is at least one incoming node that belongs to $V_{1}\cup W_{1}$, that is, there is at least one edge pointing from a node belonging to $V_{1}\cup W_{1}$ to $x_{i}$. Moreover, for any $x_{l}\in V_{1}\cup W_{1}$, there is only one outgoing node belongs to $V_{0}\cup V_{1}\cup V_{2}$, that is,  there is only one edge pointing from $x_{l}$ to a node belonging to $V_{0}\cup V_{1}\cup V_{2}$. Hence, we can get Eq. (\ref{e2}).

In a word, we have
\begin{eqnarray}
|V_{1}|+2|V_{2}|+|W_{1}|+2|W_{2}| & = & 2|U|,\label{e1}\\
|V_{0}|+|V_{1}|+|V_{2}| & \leq & |V_{1}|+|W_{1}|,\label{e2}
\end{eqnarray}
which implies that $|V_{0}|+|V_{1}|+|V_{2}|\leq 2|U|$ and then $|V|=|V_{0}|+|V_{1}|+|V_{2}|+|U|\leq3|U|$.
Therefore, in order to satisfy the controllability condition, we need to control at least $\frac{n}{3}$ nodes.
\qed

\begin{remark}
To further illustrate how Eqs. (\ref{e1}) and (\ref{e2}) are derived, we construct the following 2-2-AND-BN (see Fig. \ref{R1}),
\begin{figure}[htbp]
\begin{center}
\includegraphics[width=12cm]{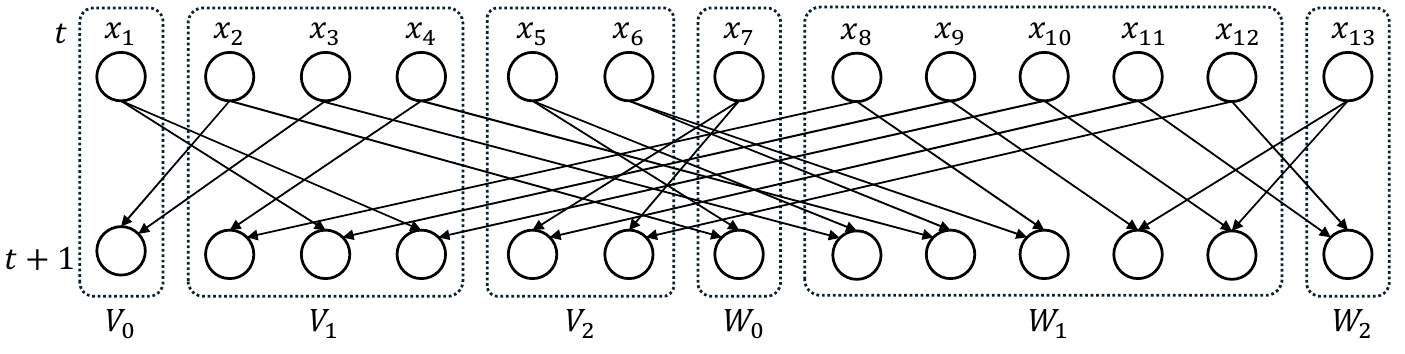}
%\caption{A 2-2-AND-BN.}
\caption{A 2-2-AND-BN, where the nodes in the top represent those at time $t$
and the nodes in the bottom represent those at time $t+1$.}
\label{R1}
\end{center}
\end{figure}
\begin{eqnarray*}
x_1(t+1) & = & x_2(t) \land x_3(t),\\
x_2(t+1) & = & x_4(t) \land x_8(t),\\
x_3(t+1) & = & x_1(t) \land x_9(t),\\
x_4(t+1) & = & x_1(t) \land x_{10}(t),\\
x_5(t+1) & = & x_7(t) \land x_{11}(t),\\
x_6(t+1) & = & x_{7}(t) \land x_{12}(t),\\
x_7(t+1) & = & x_{2}(t) \land x_{5}(t),\\
x_{8}(t+1) & = & x_{3}(t) \land x_{5}(t),\\
x_9(t+1) & = & x_{4}(t) \land x_6(t),\\
x_{10}(t+1) & = & x_{6}(t) \land x_{8}(t),\\
x_{11}(t+1) & = & x_{9}(t) \land x_{13}(t),\\
x_{12}(t+1) & = & x_{10}(t) \land x_{13}(t),\\
x_{13}(t+1) & = & x_{11}(t) \land x_{12}(t),
\end{eqnarray*}
where $V_{0}=\{x_{1}\}$, $V_{1}=\{x_{2},x_{3},x_{4}\}$, $V_{2}=\{x_{5},x_{6}\}$, $W_{0}=\{x_{7}\}$, $W_{1}=\{x_{8},x_{9},x_{10},x_{11},x_{12}\}$, and $W_{2}=\{x_{13}\}$.

It is clear that $V_{0},V_{1},V_{2},W_{0},W_{1},W_{2}$ satisfy the conditions stated in Theorem 5, where $V_{j}=\{x_{i}\mid x_{i}\notin U ~\land~ |\Gamma^{+}(x_{i})\cap U|=j~\land~ \Gamma^{+}(x_{i})\cap\{x_{i}\}=\emptyset~\land~ \Gamma^{-}(x_{i})\cap(V_{1}\cup W_{1})\neq\emptyset\}$, $W_{j} = \{x_{i}\mid x_{i}\in U ~\land~ |\Gamma^{+}(x_{i})\cap U|=j\}$,
% and $j=0,1,2$.
$j=0,1,2$, and $U=W_0 \cup W_1 \cup W_2$.

Then, for Eq. (\ref{e1}), we can see the corresponding result in Fig. \ref{R2}
because each node in $V_1 \cup W_1$ has one edge (green bold line) to $U$,
each node in $V_2 \cup W_2$ has two edges (red bold lines) to $U$,
and $U$ have edges only from $V_1 \cup V_2 \cup W_1 \cup W_2$.

\begin{figure}[htbp]
\begin{center}
\includegraphics[width=12cm]{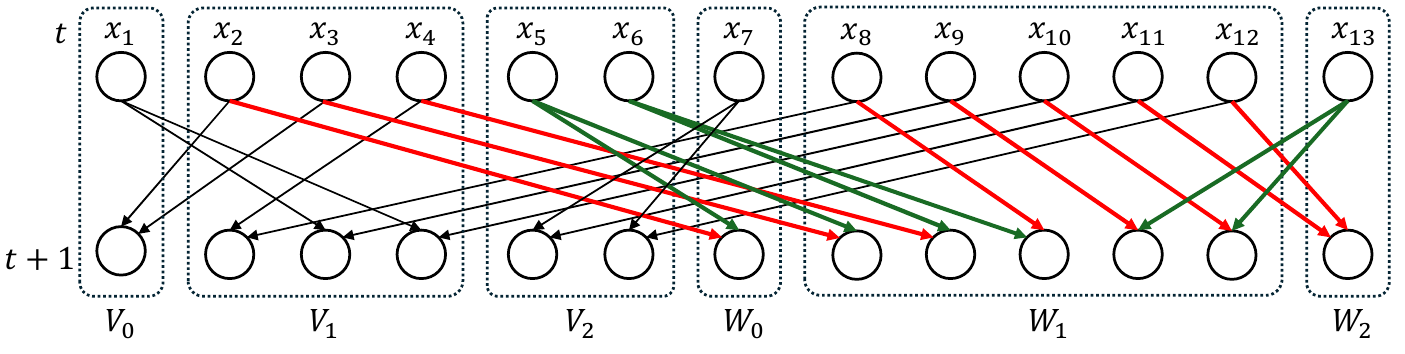}
\caption{Illustration of Eq. (1).}
\label{R2}
\end{center}
\end{figure}

In addition, for Eq. (\ref{e2}), we can see the corresponding result in Fig. \ref{R3}
because each node in $V_0 \cup V_1 \cup V_2$ has at least one edge
(blue bold line) from $V_1 \cup W_1$
(note that $\Gamma^{-}(x_i) \cap (V_1 \cup W_1) \neq \emptyset$
holds for any node $x_i \notin U$).
\begin{figure}[htbp]
\begin{center}
\includegraphics[width=12cm]{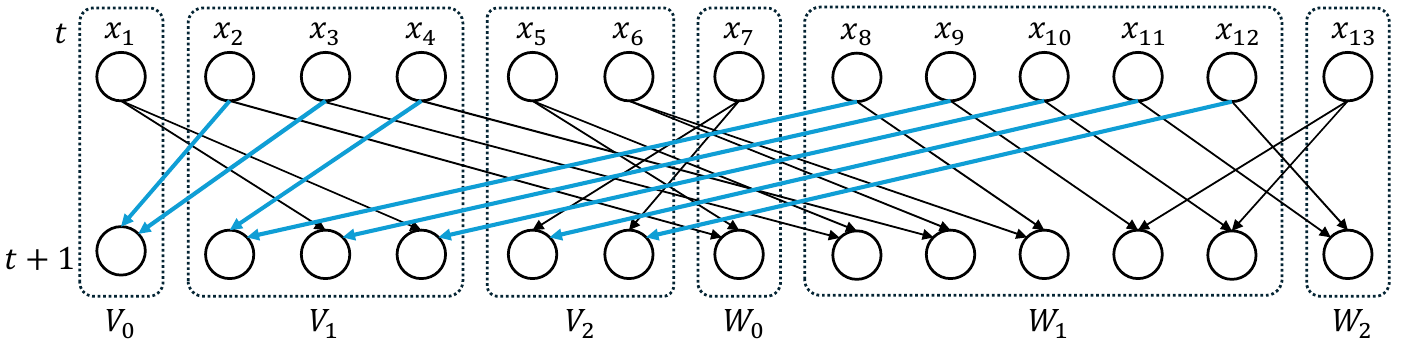}
\caption{Illustration of Eq. (2).}
\label{R3}
\end{center}
\end{figure}
\end{remark}

%\begin{remark}
%According to Eqs. (\ref{e1}) and (\ref{e2}), we have
%\begin{eqnarray*}
%2|U|&=&(|V_{1}|+|W_{1}|)+2(|V_{2}|+|W_{2}|)\\
%&\geq&(n-|U|)+2(|V_{2}|+|W_{2}|)\quad ({\rm By}~|V_{0}|+|V_{1}|+|V_{2}|=n-|U|),
%\end{eqnarray*}
%and then $|U|\geq\frac{1}{3}n+\frac{2}{3}(|V_{2}|+|W_{2}|)$. It can be seen that $|U|=\frac{1}{3}n$ only when $|V_{2}|+|W_{2}|=0$ and $|V_{1}|+|W_{1}|=n-|U|=\frac{2}{3}n$.

%Furthermore, based on $|U|=|W_{0}|+|W_{1}|+|W_{2}|$ and $|V|=|V_{0}|+|V_{1}|+|V_{2}|+|U|$, we deduce that $|U|=\frac{1}{3}n$ only when $|V_{0}|+|W_{0}|=\frac{1}{3}n,~|V_{1}|+|W_{1}|=\frac{2}{3}n,~|V_{2}|+|W_{2}|=0$ and $|V_{0}|=|W_{1}|$.
%\end{remark}

\begin{proposition}\label{proposition7}
For any integer $n = (2k-1)m+1$ with a positive integer $m>1$,
there exists a simple $k$-$k$-AND-BN
for which the size of the minimum control node set is $(k-1)m+1 = \frac{k-1}{2k-1}(n-1) + 1$.
\label{prop:k-k-and-upper}
\end{proposition}
(Proof) We construct a simple $k$-$k$-AND-BN by
\begin{eqnarray*}
x_1(t+1) & = & x_k(t) \land x_{k+1}(t) \land \cdots \land x_{2k-3}(t) \land x_{2k-2}(t) \land x_{2k}(t),\\
x_2(t+1) & = & x_k(t) \land x_{k+1}(t) \land \cdots \land x_{2k-3}(t) \land x_{2k-1}(t) \land x_{2k}(t),\\
& \vdots & \\
x_{k-1}(t+1) & = & x_k(t) \land x_{k+2}(t) \land \cdots \land x_{2k-2}(t) \land x_{2k-1}(t) \land x_{2k}(t),\\
x_{k}(t+1) & = & x_{k+1}(t) \land x_{k+2}(t) \land \cdots \land x_{2k-2}(t) \land x_{2k-1}(t) \land x_{km+k}(t),\\
x_{k+1}(t+1) & = & x_1(t) \land x_2(t) \land \cdots \land x_{k-2}(t) \land x_{k-1}(t) \land x_{k}(t),\\
x_{k+2}(t+1) & = & x_1(t) \land x_2(t) \land \cdots \land x_{k-2}(t) \land x_{k-1}(t) \land x_{k+1}(t),\\
& \vdots & \\
x_{2k}(t+1) & = & x_1(t) \land x_2(t) \land \cdots \land x_{k-2}(t) \land x_{k-1}(t) \land x_{2k-1}(t),\\
x_{2k+1}(t+1) & = & x_{km+k+1}(t) \land x_{km+k+2}(t) \land \cdots \land x_{km+2k-2}(t) \land x_{km+2k-1}(t) \land x_{2k}(t),\\
x_{2k+2}(t+1) & = & x_{km+k+1}(t) \land x_{km+k+2}(t) \land \cdots \land x_{km+2k-2}(t) \land x_{km+2k-1}(t) \land x_{2k+1}(t),\\
& \vdots & \\
x_{3k}(t+1) & = & x_{km+k+1}(t) \land x_{km+k+2}(t) \land \cdots \land x_{km+2k-2}(t) \land x_{km+2k-1}(t) \land x_{3k-1}(t),\\
& \vdots & \\
x_{km+1}(t+1) & = & x_{(2k-1)m-k+3}(t) \land x_{(2k-1)m-k+4}(t) \land \cdots \land x_{(2k-1)m}(t) \land x_{n}(t) \land x_{km}(t),\\
x_{km+2}(t+1) & = & x_{(2k-1)m-k+3}(t) \land x_{(2k-1)m-k+4}(t) \land \cdots \land x_{(2k-1)m}(t) \land x_{n}(t) \land x_{km+1}(t),\\
& \vdots & \\
x_{km+k}(t+1) & = & x_{(2k-1)m-k+3}(t) \land x_{(2k-1)m-k+4}(t) \land \cdots \land x_{(2k-1)m}(t) \land x_{n}(t) \land x_{km+k-1}(t),\\
x_{km+k+1}(t+1) & = & x_{2k+1}(t) \land x_{2k+2}(t) \land \cdots \land x_{3k-2}(t) \land x_{3k-1}(t) \land x_{3k}(t),\\
x_{km+k+2}(t+1) & = & x_{2k+1}(t) \land x_{2k+2}(t) \land \cdots \land x_{3k-2}(t) \land x_{3k-1}(t) \land x_{3k}(t),\\
& \vdots & \\
x_{km+2k-1}(t+1) & = & x_{2k+1}(t) \land x_{2k+2}(t) \land \cdots \land x_{3k-2}(t) \land x_{3k-1}(t) \land x_{3k}(t),\\
& \vdots & \\
x_{(2k-1)m-k+3}(t+1) & = & x_{km+1}(t) \land x_{km+2}(t) \land \cdots \land x_{km+k-2}(t) \land x_{km+k-1}(t) \land x_{km+k}(t),\\
x_{(2k-1)m-k+4}(t+1) & = & x_{km+1}(t) \land x_{km+2}(t) \land \cdots \land x_{km+k-2}(t) \land x_{km+k-1}(t) \land x_{km+k}(t),\\
& \vdots & \\
x_{n}(t+1) & = & x_{km+1}(t) \land x_{km+2}(t) \land \cdots \land x_{km+k-2}(t) \land x_{km+k-1}(t) \land x_{km+k}(t)
\end{eqnarray*}
with letting $x_1,x_2,\ldots,x_{k}$, and $x_{km+k+i}$ ($i=1,\dots,(k-1)m-(k-1)$)
be control nodes.
By using control signals,
we let $x_1(t)=x_{2}(t)=\cdots=x_{k-1}(t)=1$ and $x_{km+k+i}(t)=1$ ($i=1,\dots,(k-1)m-(k-1)$)
for all $t=1,\ldots,km$,
$x_1(km+1)=x^T_1,~x_2(km+1)=x^T_2,\ldots,x_{k-1}(km+1)=x^T_{k-1}$.
$x_{km+k+i}(km+1)=x_{km+k+i}^T$ ($i=1,\dots,(k-1)m-(k-1)$),
and $x_k(t) = x^T_{km+k+1-t}$ for $t=1,\ldots,km+1$.
Then, it is straightforward to see that $\xvec(km+1)=\xvec^T$.
\qed

\bigskip

\begin{example}
According to Proposition \ref{prop:k-k-and-upper}, we can construct a 3-3-AND-BN (see Fig. \ref{fig:3-3}), where $m=2$ as follows:
\begin{eqnarray*}
x_1(t+1) & = & x_3(t) \land x_4(t) \land x_{6}(t),\\
x_2(t+1) & = & x_3(t) \land x_5(t) \land x_{6}(t),\\
x_3(t+1) & = & x_4(t) \land x_5(t) \land x_{9}(t),\\
x_4(t+1) & = & x_1(t) \land x_2(t) \land x_3(t),\\
x_5(t+1) & = & x_1(t) \land x_2(t) \land x_4(t),\\
x_6(t+1) & = & x_1(t) \land x_2(t) \land x_5(t),\\
x_7(t+1) & = & x_{10}(t) \land x_{11}(t) \land x_6(t),\\
x_8(t+1) & = & x_{10}(t) \land x_{11}(t) \land x_7(t),\\
x_9(t+1) & = & x_{10}(t) \land x_{11}(t) \land x_8(t),\\
x_{10}(t+1) & = & x_7(t) \land x_8(t) \land x_9(t),\\
x_{11}(t+1) & = & x_7(t) \land x_8(t) \land x_9(t).
\end{eqnarray*}
\begin{figure}[th]
\begin{center}
\includegraphics[width=10cm]{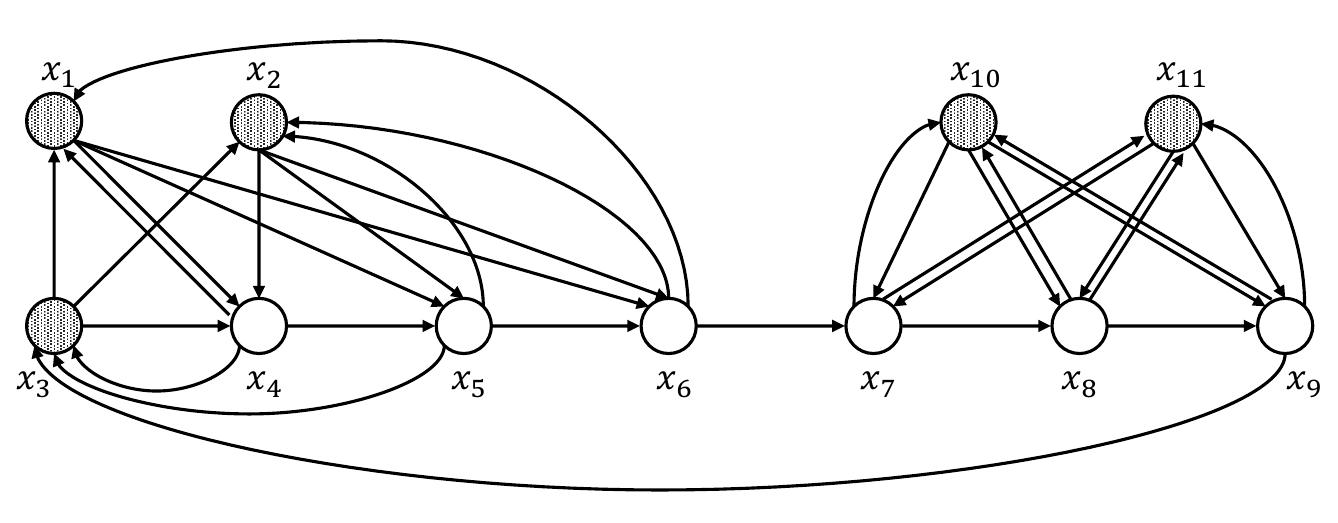}
\caption{A 3-3-AND-BN, where $x_{1},x_{2},x_{3},x_{10},x_{11}$ are control nodes.}
\label{fig:3-3}
\end{center}
\end{figure}
In this case, let $U=\{x_{1},x_{2},x_{3},x_{10},x_{11}\}$. By using control signals,
we let $x_1(t)=x_{2}(t)=1$ and $x_{10}(t)=x_{11}(t)=1$
for all $t=1,\ldots,6$,
$x_1(7)=x^T_1,~x_2(7)=x^T_2$.
$x_{10}(7)=x_{10}^T,~x_{11}(7)=x_{11}^T$,
and $x_3(t) = x^T_{10-t}$ for $t=1,\ldots,7$.
Then, it is straightforward to see that $\xvec(7)=\xvec^T$.

Suppose $\xvec^0=[0,0,0,0,0,0,0,0,0,0,0]$ and $\xvec^T=[1,0,0,1,1,0,1,0,0,1,1]$. By using the above control signals, we have the following.
\begin{center}
\begin{tabular}{c|lllllllllll}
\hline
$\xvec^0$ ($t=0$) & 0 & 0 & 0 & 0 & 0 & 0 & 0 & 0 & 0 & 0 & 0\\
$\xvec(1)$ & 1 & 1 & 0 & 0 & 0 & 0 & 0 & 0 & 0 & 1 & 1\\
$\xvec(2)$ & 1 & 1 & 0 & 0 & 0 & 0 & 0 & 0 & 0 & 1 & 1\\
$\xvec(3)$ & 1 & 1 & 1 & 0 & 0 & 0 & 0 & 0 & 0 & 1 & 1\\
$\xvec(4)$ & 1 & 1 & 0 & 1 & 0 & 0 & 0 & 0 & 0 & 1 & 1\\
$\xvec(5)$ & 1 & 1 & 1 & 0 & 1 & 0 & 0 & 0 & 0 & 1 & 1\\
$\xvec(6)$ & 1 & 1 & 1 & 1 & 0 & 1 & 0 & 0 & 0 & 1 & 1\\
$\xvec^T$ ($t=7$) & 1 & 0 & 0 & 1 & 1 & 0 & 1 & 0 & 0 & 1 & 1\\
\hline
\end{tabular}
\end{center}
\end{example}

\begin{theorem}\label{thm:k-k-AND}
For any simple $k$-$k$-AND-BN, the size of the control node set is at least $\frac{k-1}{2k-1}n$.
\end{theorem}
(Proof)
Note that for any node $x_{i}$ in a simple $k$-$k$-AND-BN, we have $|\Gamma^{+}(x_{i})|=|\Gamma^{-}(x_{i})|=k$.

Similarly, if a simple $k$-$k$-AND-BN is controllable, any node $x_{i}\notin U$ will not encounter the case that $x_{i}\in\Gamma^{+}(x_{i})$.

Now, we divide any node $x_{i}\notin U$ in a controllable simple $k$-$k$-AND-BN into $k+1$ disjoint sets  $V_0,V_{1},\ldots,V_{k}$, where
\begin{eqnarray*}
V_{0} & = & \{x_{i}\mid x_{i}\notin U ~\land~ \Gamma^{+}(x_{i})\cap U=\emptyset~\land~ \Gamma^{+}(x_{i})\cap\{x_{i}\}=\emptyset\},\\
V_{1} & = & \{x_{i}\mid x_{i}\notin U ~\land~ |\Gamma^{+}(x_{i})\cap U|=1 ~\land~ \Gamma^{+}(x_{i})\cap\{x_{i}\}=\emptyset\},\\
&\vdots&\\
V_{k} & = & \{x_{i}\mid x_{i}\notin U ~\land~|\Gamma^{+}(x_{i})\cap U|=k\},
\end{eqnarray*}
and we divide any node $x_{j}\in U$ in a controllable simple $k$-$k$-AND-BN into $k+1$ disjoint sets  $W_0,W_1,\ldots,W_k$, where
\begin{eqnarray*}
W_{0} & = & \{x_{j}\mid x_{j}\in U ~\land~ \Gamma^{+}(x_{j})\cap U=\emptyset\},\\
W_{1} & = & \{x_{j}\mid x_{j}\in U ~\land~ |\Gamma^{+}(x_{j})\cap U|=1\},\\
&\vdots&\\
W_{k} & = & \{x_{j}\mid x_{j}\in U ~\land~ |\Gamma^{+}(x_{j})\cap U|=k\}.
\end{eqnarray*}
Notably, $|V|=\sum_{p=0}^{k}|V_{p}|+|U|=\sum_{p=0}^{k}|V_{p}|+\sum_{p=0}^{k}|W_{p}|$.

Then, it is seen that if a simple $k$-$k$-AND-BN is controllable, then for any node $x_{i}\notin U$, we have $\Gamma^{-}(x_{i})\cap(V_{k-1}\cup W_{k-1})\neq\emptyset$. Suppose there is a node $x_{i}\notin U$, where $\Gamma^{-}(x_{i})\cap V_{k-1}=\emptyset$ and $\Gamma^{-}(x_{i})\cap W_{k-1}=\emptyset$. We assume without loss of generality $x_{i}(t+1)=x_{i_{1}}(t)\wedge x_{i_{2}}(t)\wedge\cdots\wedge x_{i_{k}}(t)$, where $x_{i_{1}},x_{i_{2}},\ldots x_{i_{k}}\notin V_{k-1}\cup W_{k-1}$. Clearly, $x_{i_{1}},x_{i_{2}},\ldots x_{i_{k}}\notin V_{k}\cup W_{k}$, because for any $x_{l}\in V_{k}\cup W_{k}$,  it is definitely not an incoming node for any $x_{i}\notin U$. Then we assume without loss of generality $x_{l_{1}}\in\Gamma^{+}(x_{i_{1}})\setminus\{x_{i}\},x_{l_{2}}\in\Gamma^{+}(x_{i_{2}})\setminus\{x_{i}\},\ldots, x_{l_{k}}\in\Gamma^{+}(x_{i_{k}})\setminus\{x_{i}\}$ and $x_{l_{1}},x_{l_{2}},\ldots,x_{l_{k}}\notin U$.
Here, some of $x_{l_1},x_{l_{2}},\ldots,x_{l_k}$ can be the same, and $x_{i_{p}},~p=1,2,\ldots,k$ can belong to $U$ or not belong to $U$. Then for any initial state $\xvec^0$, the target state of the form $x_{i}=0,x_{l_{1}}=x_{l_{2}}=\cdots=x_{l_{k}}=1$ cannot be reached, which contradicts the fact that the simple $k$-$k$-AND-BN is controllable.

Therefore, if a simple $k$-$k$-AND-BN is controllable, then for any $x_{i}\notin U$, there are $k+1$ cases:
\begin{itemize}
  \item $|\Gamma^{+}(x_{i})\cap U|=k$ and $\Gamma^{-}(x_{i})\cap(V_{k-1}\cup W_{k-1})\neq\emptyset$;
  \item $|\Gamma^{+}(x_{i})\cap U|=k-1$, $\Gamma^{+}(x_{i})\cap\{x_{i}\}=\emptyset$ and $\Gamma^{-}(x_{i})\cap(V_{k-1}\cup W_{k-1})\neq\emptyset$;
  \item $\ldots$
  \item $\Gamma^{+}(x_{i})\cap U=\emptyset$, $\Gamma^{+}(x_{i})\cap\{x_{i}\}=\emptyset$ and $\Gamma^{-}(x_{i})\cap(V_{k-1}\cup W_{k-1})\neq\emptyset$.
\end{itemize}

Based on the above analysis, for any node $x_{l}\in V_{p}\cup W_{p},~p=1,2,\ldots,k$, there are $p$ outgoing nodes that belongs to $U$, that is, there are $p$ edges pointing from $x_{l}$ to the control nodes. It is known that the total number of in-degrees of control nodes is $k|U|$, and for any node $x_{l}\in V_{0}\cup W_{0}$, there is no outgoing node belonging to $U$. Hence, we can get Eq. (\ref{e3}).

On the other hand, if a simple $k$-$k$-AND-BN is controllable, then for any node $x_{i}\in \cup_{p=0}^{k}V_{p}$, there is at least one incoming node that belongs to $V_{k-1}\cup W_{k-1}$, that is, there is at least one edge pointing from a node belonging to $V_{k-1}\cup W_{k-1}$ to $x_{i}$. Moreover, for any $x_{l}\in V_{k-1}\cup W_{k-1}$, there is only one outgoing node belongs to $\cup_{p=0}^{k}V_{p}$, that is,  there is only one edge pointing from $x_{l}$ to a node belonging to $\cup_{p=0}^{k}V_{p}$. Hence, we can get Eq. (\ref{e4}).

In a word, we have
\begin{eqnarray}
|V_{1}|+2|V_{2}|+\cdots+k|V_{k}|+|W_{1}|+2|W_{2}|+\cdots+k|W_{k}| & = & k|U|,\label{e3}\\
|V_{0}|+|V_{1}|+\cdots+|V_{k-1}|+|V_{k}| & \leq & |V_{k-1}|+|W_{k-1}|,\label{e4}
\end{eqnarray}
which implies that
\begin{eqnarray*}
k|U|&\geq&(k-1)|V_{k-1}|+(k-1)|W_{k-1}|\\
&\geq&(k-1)(|V_{0}|+|V_{1}|+\cdots+|V_{k-1}|+|V_{k}|)
\end{eqnarray*}
and then
\begin{eqnarray*}
|V|&=&|V_{0}|+|V_{1}|+\cdots+|V_{k}|+|U|\\
&\leq&\frac{k}{k-1}|U|+|U|\\
&=&\frac{2k-1}{k-1}|U|.
\end{eqnarray*}
Therefore, in order to satisfy the controllability condition, we need to control at least $\frac{k-1}{2k-1}n$ nodes.
\qed

\bigskip

\begin{remark}
According to Eqs. (\ref{e3}) and (\ref{e4}), we have
\begin{eqnarray*}
k|U|&=&(|V_{1}|+|W_{1}|)+2(|V_{2}|+|W_{2}|)+\cdots+(k-1)(|V_{k-1}|+|W_{k-1}|)+k(|V_{k}|+|W_{k}|)\\
&\geq&(|V_{1}|+|W_{1}|)+\cdots+(k-2)(|V_{k-2}|+|W_{k-2}|)+(k-1)(n-|U|)+k(|V_{k}|+|W_{k}|),
\end{eqnarray*}
and then $|U|\geq\frac{k-1}{2k-1}n+\frac{1}{2k-1}[(|V_{1}|+|W_{1}|)+\cdots+(k-2)(|V_{k-2}|+|W_{k-2}|)+k(|V_{k}|+|W_{k}|)]$. It can be seen that $|U|=\frac{k-1}{2k-1}n$ only when $|V_{1}|+|W_{1}|=0,\ldots,|V_{k-2}|+|W_{k-2}|=0,|V_{k}|+|W_{k}|=0$ and $|V_{k-1}|+|W_{k-1}|=n-|U|=\frac{k}{2k-1}n$.

Furthermore, based on $|U|=|W_{0}|+|W_{1}|+\cdots+|W_{k}|$ and $|V|=|V_{0}|+|V_{1}|+\cdots+|V_{k}|+|U|$, we deduce that $|U|=\frac{k-1}{2k-1}n$ only when $|V_{0}|+|W_{0}|=\frac{k-1}{2k-1}n,|V_{1}|+|W_{1}|=0,\ldots,|V_{k-2}|+|W_{k-2}|=0,|V_{k-1}|+|W_{k-1}|=\frac{k}{2k-1}n,|V_{k}|+|W_{k}|=0$ and $|V_{0}|=|W_{k-1}|$.
\end{remark}

\begin{remark}
The above results for simple $k$-$k$-AND-BN ($k\geq2$) also apply to simple $k$-$k$-OR-BN ($k\geq2$). In other words, for any simple $k$-$k$-OR-BN, the size of the control node set is at least $\frac{k-1}{2k-1}n$.
\end{remark}
\section{Boolean networks consisting of AND/OR functions and negations}
Next, we consider the occurrence of literal $\lnon{x}$ in AND functions, and generalize the simple 2-2-AND-BN to the 2-2-AND-BN with negation, that is, each function in this BN is of the form $x \land y$, $x \land \lnon{y}$, $\lnon{x} \land y$, or $\lnon{x} \land \lnon{y}$. Specifically, when each function in a 2-2-AND-BN with negation is of the form $x \land y$, then the BN is a simple 2-2-AND-BN.

Let $LIT^{-}(x_{i})$ denote the set of incoming literals. Define $M_{0}$ as the set of nodes where only the literal $x_ {i}$ or $\lnon{x_{i}}$ appears in the 2-2-AND-BN with negation. Define $M_{1}$ as the set of nodes where
both literals $x_{i}$ and $\lnon{x_{i}}$ appear in the 2-2-AND-BN with negation. In other words, since the outdegree of each node in the 2-2-AND-BN with negation is 2, for any node $x_{i}\in M_{1}$, the corresponding literals $x_{i}$ and $\lnon{x_{i}}$ appear once in the 2-2-AND-BN with negation, respectively. Then $M_{0}\cup M_{1}=V$. For example, if $x_{1}(t+1)=x_{2}(t) \land \lnon{x_{3}(t)},~x_{2}(t+1)=x_{1}(t) \land \lnon{x_{2}(t)},~x_{3}(t+1)=x_{1}(t) \land \lnon{x_{3}(t)}$, then $LIT^{-}(x_{1})=\{x_{2},\lnon{x_{3}}\},~LIT^{-}(x_{2})=\{x_{1},\lnon{x_{2}}\},~LIT^{-}(x_{3})=\{x_{1},\lnon{x_{3}}\}$ and $M_{0}=\{x_{1},x_{3}\},~M_{1}=\{x_{2}\}$.

First, we consider a special case where each node $x_{i}$ in the 2-2-AND-BN with negation has two types of literals, $x_{i}$ and $\overline{x_{i}}$, i.e., $M_{0}=\emptyset$ and $M_{1}=V$.

\begin{example}\label{example5}
According to Proposition \ref{prop:2-2-and-upper}, we construct a 2-2-AND-BN with negation, where $M_{1}=V$ and $m=2$ as follows:
\begin{eqnarray*}
x_1(t+1) & = & \lnon{x_2(t)} \land x_4(t),\\
x_2(t+1) & = & x_3(t) \land \lnon{x_{6}(t)},\\
x_3(t+1) & = & \lnon{x_1(t)} \land x_{2}(t),\\
x_4(t+1) & = & x_1(t) \land \lnon{x_3(t)},\\
x_5(t+1) & = & \lnon{x_4(t)} \land x_7(t),\\
x_6(t+1) & = & x_5(t) \land \lnon{x_7(t)},\\
x_7(t+1) & = & \lnon{x_{5}(t)} \land x_6(t).
\end{eqnarray*}
In this case, if $U$ is still set to $\{x_{1},x_{2},x_{7}\}$ according to Proposition \ref{prop:2-2-and-upper}, then no initial state can reach the target state of the form $x_{3}=x_{4}=1$ and/or $x_{5}=x_{6}=1$, because $\lnon{x_{1}}\in LIT^{-}(x_{3}),~x_{1}\in LIT^{-}(x_{4})$ and $x_{7}\in LIT^{-}(x_{5}),~\lnon{x_{7}}\in LIT^{-}(x_{6})$.

Here, we note that if the situation that $\lnon{x_{1}}\in LIT^{-}(x_{3}),~x_{1}\in LIT^{-}(x_{4}),~x_{3},x_{4}\notin U$ occurs, then the 2-2-AND-BN with negation cannot be controllable. Furthermore, in order to avoid the above situation, we found that $|U|$ must be greater than 3 in this case.

Now let $U=\{x_{3},x_{4},x_{5},x_{7}\}$. By using control signals, we let $x_{3}(t)=x_{4}(t)=x_{5}(t)=1$ for $t=1,2,3$, $\lnon{x_{7}(1)}=x_{1}^{T}$, $x_{7}(2)=x_{2}^{T}$, $\lnon{x_{7}(3)}=x_{6}^{T}$, and $x_{i}(4)=x_{i}^{T},~i=3,4,5,7$. Then, it is straightforward to see that $\xvec(4)=\xvec^T$ and $U=\{x_{3},x_{4},x_{5},x_{7}\}$ is the minimum control node set in this case.
\end{example}

\begin{theorem}\label{thm:2-2-AND-negation}
For any 2-2-AND-BN with negation, the size of the
control node set is at least
\begin{eqnarray*}
\max\left(\frac{1}{3}n,\frac{1}{2}|M_{1}|\right),
\end{eqnarray*}
where $M_{1}$ is the set of nodes where both literals $x_{i}$ and $\lnon{x_{i}}$ appear in the 2-2-AND-BN with negation.
\end{theorem}
(Proof) First, the relations given by Eqs. (\ref{e1}) and (\ref{e2}) in Theorem \ref{thm:2-2-and} remain valid in this case.
Then we note that for any node $x_{i}\in M_{1}$, $|\Gamma^{+}(x_{i})\cap U|\geq1$. Suppose there is a node  $x_{i}\in M_{1}$, where $|\Gamma^{+}(x_{i})\cap U|=0$. We assume without loss of generality $x_{i_{1}}(t+1)=x_{i}(t)\land x_{k_{1}}(t)$ and  $x_{i_{2}}(t+1)=\lnon{x_{i}(t)}\land x_{k_{2}}(t)$. Here, $x_{i_{1}},x_{i_{2}}\notin U$, because $|\Gamma^{+}(x_{i})\cap U|=0$. Then for any initial state $\xvec^0$, the target state of the form $x_{i_{1}}=x_{i_{2}}=1$ cannot be reached, which contradicts the fact that the 2-2-AND-BN with negation is controllable.

According to the above analysis, in order to ensure that the 2-2-AND-BN with negation is controllable, the following two conditions need to be satisfied,
\begin{itemize}
  \item[(i)] $|V_{1}|+|W_{1}|\geq n-|U|$;
  \item[(ii)] for any node $x_{i}\in M_{1}$, $|\Gamma^{+}(x_{i})\cap U|\geq1$.
\end{itemize}
Notably, for any node $x_{i}\in M_{1}$, it is possible that $x_{i}$ belongs to $V_{1}\cup W_{1}$.

Our goal is to find the minimum control node set $U$ that satisfies the above two conditions. Therefore, according to Eq. (\ref{e1}), for any node $x_{i}\in M_{1}$, let $|\Gamma^{+}(x_{i})\cap U|=1$. Then there are two cases as follows:
\begin{itemize}
  \item When $n-|U|>|M_{1}|$, we can minimize $2|U|=(|V_{1}|+|W_{1}|)+2(|V_{2}|+|W_{2}|)$ to $n-|U|$, where $M_{1}\subseteq V_{1}\cup W_{1},~|V_{1}|+|W_{1}|=n-|U|$ and $|V_{2}|+|W_{2}|=0$, while satisfying the above two conditions.
  \item When $0<n-|U|\leq|M_{1}|$, under the premise that the above two conditions are satisfied, the minimum value of $2|U|=(|V_{1}|+|W_{1}|)+2(|V_{2}|+|W_{2}|)$ is $|M_{1}|$, where $|V_{1}|+|W_{1}|=|M_{1}|\geq n-|U|$ and $|V_{2}|+|W_{2}|=0$.
\end{itemize}
In a word, according to Eq. (\ref{e1}), we have
\begin{eqnarray*}
2|U| \geq \left\{
\begin{array}{ll}
n-|U|, & n-|U|>|M_{1}|,\\
|M_{1}|, & 0<n-|U|\leq|M_{1}|.
\end{array}
\right.
\end{eqnarray*}
that is,
$2|U|\geq\max(n-|U|,|M_{1}|)$.

Therefore, in order to satisfy the controllability condition, we need to control at least $\max\left(\frac{1}{3}n,\frac{1}{2}|M_{1}|\right)$ nodes.
\qed

\bigskip

\begin{remark}
According to Theorem \ref{thm:2-2-AND-negation}, it is noted that for any 2-2-AND-BN with negation, the set of control node set is at least $\frac{1}{2}|M_{1}|$ when $|M_{1}|>\frac{2}{3}n$, which is consistent with the fact that the size of the minimum control node set in Example \ref{example5} is 4 ($|M_{1}|=n=7$).
On the other hand, the set of control node set is at least $\frac{1}{3}n$ when $|M_{1}|\leq\frac{2}{3}n$, which is consistent with the conclusion of Theorem \ref{thm:2-2-and} ($M_{1}=\emptyset$). In other words, Theorem \ref{thm:2-2-and} is a special case of Theorem \ref{thm:2-2-AND-negation}, which is also natural due to the fact that the simple 2-2-AND BN is a special case of the 2-2-AND-BN with negation. Additionally, when $|M_{1}|>\frac{2}{3}n$, the general lower bound given in Theorem \ref{thm:2-2-AND-negation} is larger than the general lower bound given in Theorem \ref{thm:2-2-and}.
\end{remark}

Then we extend Theorem \ref{thm:2-2-AND-negation} to $k$-$k$-AND-BNs with negation. Specifically, when each function in a $k$-$k$-AND-BN with negation is of the form $x_1 \land x_2 \land \cdots \land x_k$, then the BN is a simple $k$-$k$-AND-BN.

Define $M_{j},~j=0,1,\ldots,\lfloor\frac{k}{2}\rfloor$ as the set of nodes where the literal $x_{i}$ or $\lnon{x_{i}}$ ($i=1,2,\ldots,n$) appears $j$ times in the $k$-$k$-AND-BN with negation. Since the outdegree of each node in the $k$-$k$-AND-BN with negation is $k$, for any node $x_{i}\in M_{j}$, there are two cases, one is that the literal $x_{i}$ occurs $j$ times and the literal $\lnon{x_{i}}$ occurs $k-j$ times; the other is that the literal $x_{i}$ occurs $k-j$ times and the literal $\lnon{x_{i}}$ occurs $j$ times. For example, if $x_{1}(t+1)=\lnon{x_{1}(t)}\land x_{2}(t) \land \lnon{x_{3}(t)},~x_{2}(t+1)=x_{1}(t)\land \lnon{x_{2}(t)} \land \lnon{x_{3}(t)},~x_{3}(t+1)=x_{1}(t)\land \lnon{x_{2}(t)} \land \lnon{x_{3}(t)}$, then $M_{0}=\{x_{3}\}$ and $M_{1}=\{x_{1},x_{2}\}$.

\begin{theorem}\label{thm:k-k-AND-negation}
For any $k$-$k$-AND-BN with negation, the size of the control node set is at least
\begin{eqnarray*}
\max\left\{\frac{k-l}{2k-l}n+\frac{1}{2k-l}\left(\sum_{j=1}^{l-2}j\cdot|M_{j}|\right)+\frac{l-1}{2k-l}\left(\sum_{j=l-1}^{\lfloor\frac{k}{2}\rfloor}|M_{j}|\right),~l=1,2,\ldots,\lfloor\frac{k}{2}\rfloor+1\right\},
\end{eqnarray*}
where $M_{j},~j=0,1,\ldots,\lfloor\frac{k}{2}\rfloor$ is the set of nodes where the literal $x_{i}$ or $\lnon{x_{i}}$ ($i=1,2,\ldots,n$) appears $j$ times in the $k$-$k$-AND-BN with negation, and for any $k_{1}<k_{2}$, $\sum_{j=k_{2}}^{k_{1}}|M_{j}|=0$.
\end{theorem}

(Proof)
First, the relations given by Eqs. (\ref{e3}) and (\ref{e4}) in Theorem \ref{thm:k-k-AND} remain valid in this case.

We note that for any node $x_{i}\in M_{1}$, $|\Gamma^{+}(x_{i})\cap U|\geq1$. Suppose there is a node $x_{i}\in M_{1}$, where $|\Gamma^{+}(x_{i})\cap U|=0$. We assume without loss of generality $x_{i_{1}}(t+1)=x_{i}(t)\land x_{l_{1}}(t)\land \cdots\land x_{l_{k-1}}(t)$ and  $x_{i_{2}}(t+1)=\lnon{x_{i}(t)}\land x_{l_{k}}(t)\land \cdots\land x_{l_{2k-1}}(t)$. Here, $x_{i_{1}},x_{i_{2}}\notin U$, because $|\Gamma^{+}(x_{i})\cap U|=0$. Then for any initial state $\xvec^0$, the target state of the form $x_{i_{1}}=x_{i_{2}}=1$ cannot be reached, which contradicts the fact that the $k$-$k$-AND-BN with negation is controllable. Similarly, we deduce that for any node $x_{i}\in M_{j},~j=1,2,\ldots,\lfloor\frac{k}{2}\rfloor$, $|\Gamma^{+}(x_{i})\cap U|\geq j$.

\begin{figure}[th]
\begin{center}
\includegraphics[width=10cm]{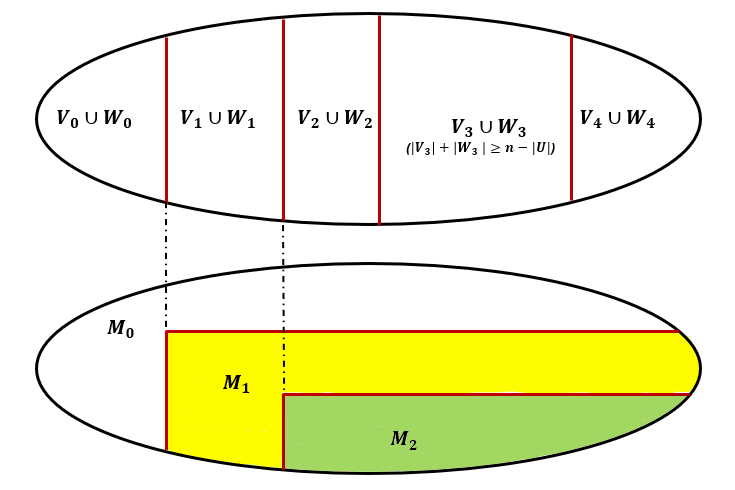}
\caption{For a controllable 4-4-AND-BN with negation, it can be seen that $|V_{3}|+|W_{3}|\geq n-|U|,~M_{1}\cap(V_{0}\cup W_{0})=\emptyset$ and $M_{2}\cap(V_{0}\cup W_{0}\cup V_{1}\cup W_{1})=\emptyset$.}
\label{fig5}
\end{center}
\end{figure}

According to the above analysis, in order to ensure that the $k$-$k$-AND-BN with negation is controllable, the following two conditions need to be satisfied,
\begin{itemize}
  \item[(i)] $|V_{k-1}|+|W_{k-1}|\geq n-|U|$;
  \item[(ii)] for any node $x_{i}\in M_{j},~j=1,2,\ldots,\lfloor\frac{k}{2}\rfloor$, $|\Gamma^{+}(x_{i})\cap U|\geq j$.
\end{itemize}
Notably, for any node $x_{i}\in M_{j},~j=1,2,\ldots,\lfloor\frac{k}{2}\rfloor$, it is possible that $x_{i}$ belongs to $V_{k-1}\cup W_{k-1}$ (See Fig. \ref{fig5}).

Our goal is to find the minimum control node set $U$ that satisfies the above two conditions. In other words, we want to minimize the value of $k|U|=(|V_{1}|+|W_{1}|)+2(|V_{2}|+|W_{2}|)+\cdots+k(|V_{k}|+|W_{k}|)$ subject to $|V_{k-1}|+|W_{k-1}|\geq n-|U|$ and $|\Gamma^{+}(x_{i})\cap U|\geq j$ for any $x_{i}\in M_{j},~j=1,2,\ldots,\lfloor\frac{k}{2}\rfloor$, so we have
\begin{eqnarray*}
&&k|U|\\ & = & (|V_{1}|+|W_{1}|)+\cdots+\lfloor\frac{k}{2}\rfloor(|V_{\lfloor\frac{k}{2}\rfloor}|+|W_{\lfloor\frac{k}{2}\rfloor}|)+\cdots+(k-1)(|V_{k-1}|+|W_{k-1}|)+k(|V_{k}|+|W_{k}|),\\
& = & (|V_{1}|+|W_{1}|)+\cdots+\lfloor\frac{k}{2}\rfloor(|V_{\lfloor\frac{k}{2}\rfloor}|+|W_{\lfloor\frac{k}{2}\rfloor}|)+\cdots+(k-1)(n-|U|)+k(|V_{k}|+|W_{k}|),\\
 & = & (|V_{1}|+|W_{1}|)+\cdots+\lfloor\frac{k}{2}\rfloor(|V_{\lfloor\frac{k}{2}\rfloor}|+|W_{\lfloor\frac{k}{2}\rfloor}|)+0+\cdots+0+(k-1)(n-|U|)+0,\\
 &=& (|V_{1}|+|W_{1}|)+\cdots+\lfloor\frac{k}{2}\rfloor(|V_{\lfloor\frac{k}{2}\rfloor}|+|W_{\lfloor\frac{k}{2}\rfloor}|)+(k-1)(n-|U|).
\end{eqnarray*}
Here, because we want $(|V_{1}|+|W_{1}|)+2(|V_{2}|+|W_{2}|)+\cdots+(k-2)(|V_{k-2}|+|W_{k-2}|)+k(|V_{k}|+|W_{k}|)$ to be as small as possible, we let $|V_{k-1}|+|W_{k-1}|=n-|U|$ and set $|\Gamma^{+}(x_{i})\cap U|=j~{\rm or}~k-1$, for any node $x_{i}\in M_{j}$ ($j=1,\ldots,\lfloor\frac{k}{2}\rfloor$), so that
the terms $(\lfloor\frac{k}{2}\rfloor+1)(|V_{\lfloor\frac{k}{2}\rfloor+1}|+|W_{\lfloor\frac{k}{2}\rfloor+1}|),\ldots,(k-2)(|V_{k-2}|+|W_{k-2}|),k(|V_{k}|+|W_{k}|)$ all take the value of 0. Now, according to the fact that $\left(\cup_{j=1}^{\lfloor\frac{k}{2}\rfloor}M_{j}\right)\cap(V_{k-1}\cup W_{k-1})\neq\emptyset$, there are several cases as follows:
\begin{itemize}
  \item When $n-|U|>\sum_{j=1}^{\lfloor\frac{k}{2}\rfloor}|M_{j}|$, we can minimize $k|U|=(|V_{1}|+|W_{1}|)+\cdots+(k-1)(|V_{k-1}|+|W_{k-1}|)+k(|V_{k}|+|W_{k}|)$ to $(k-1)(n-|U|)$, where $\cup_{j=1}^{\lfloor\frac{k}{2}\rfloor}M_{j}\subseteq V_{k-1}\cup W_{k-1},~|V_{k-1}|+|W_{k-1}|=n-|U|$ and $|V_{1}|+|W_{1}|=\cdots=|V_{k-2}|+|W_{k-2}|=|V_{k}|+|W_{k}|=0$, while satisfying the above two conditions.
  \item When $\sum_{j=2}^{\lfloor\frac{k}{2}\rfloor}|M_{j}|<n-|U|\leq\sum_{j=1}^{\lfloor\frac{k}{2}\rfloor}|M_{j}|$, under the premise that the above two conditions are satisfied, the minimum value of $k|U|=(|V_{1}|+|W_{1}|)+\cdots+(k-1)(|V_{k-1}|+|W_{k-1}|)+k(|V_{k}|+|W_{k}|)$ is $(k-1)(n-|U|)+\left[\sum_{j=1}^{\lfloor\frac{k}{2}\rfloor}|M_{j}|-(n-|U|)\right]$, where $\cup_{j=2}^{\lfloor\frac{k}{2}\rfloor}M_{j}\subseteq V_{k-1}\cup W_{k-1},~|V_{k-1}|+|W_{k-1}|=n-|U|,~|V_{1}|+|W_{1}|=\sum_{j=1}^{\lfloor\frac{k}{2}\rfloor}|M_{j}|-(n-|U|)$ and $|V_{2}|+|W_{2}|=\cdots=|V_{k-2}|+|W_{k-2}|=|V_{k}|+|W_{k}|=0$.
  \item When $\sum_{j=3}^{\lfloor\frac{k}{2}\rfloor}|M_{j}|<n-|U|\leq\sum_{j=2}^{\lfloor\frac{k}{2}\rfloor}|M_{j}|$, under the premise that the above two conditions are satisfied, the minimum value of $k|U|=(|V_{1}|+|W_{1}|)+\cdots+(k-1)(|V_{k-1}|+|W_{k-1}|)+k(|V_{k}|+|W_{k}|)$ is $(k-1)(n-|U|)+|M_{1}|+2\left[\sum_{j=2}^{\lfloor\frac{k}{2}\rfloor}|M_{j}|-(n-|U|)\right]$, where $\cup_{j=3}^{\lfloor\frac{k}{2}\rfloor}M_{j}\subseteq V_{k-1}\cup W_{k-1},~|V_{k-1}|+|W_{k-1}|=n-|U|,~V_{1}\cup W_{1}=M_{1},~|V_{2}|+|W_{2}|=\sum_{j=2}^{\lfloor\frac{k}{2}\rfloor}|M_{j}|-(n-|U|)$ and $|V_{3}|+|W_{3}|=\cdots=|V_{k-2}|+|W_{k-2}|=|V_{k}|+|W_{k}|=0$.
  \item $\ldots$
  \item When $|M_{\lfloor\frac{k}{2}\rfloor}|<n-|U|\leq|M_{\lfloor\frac{k}{2}\rfloor-1}|+|M_{\lfloor\frac{k}{2}\rfloor}|$, under the premise that the above two conditions are satisfied, the minimum value of $k|U|=(|V_{1}|+|W_{1}|)+\cdots+(k-1)(|V_{k-1}|+|W_{k-1}|)+k(|V_{k}|+|W_{k}|)$ is $(k-1)(n-|U|)+\sum_{j=1}^{\lfloor\frac{k}{2}\rfloor-2}j\cdot|M_{j}|+(\lfloor\frac{k}{2}\rfloor-1)\left[|M_{\lfloor\frac{k}{2}\rfloor-1}|+|M_{\lfloor\frac{k}{2}\rfloor}|-(n-|U|)\right]$, where $M_{\lfloor\frac{k}{2}\rfloor}\subseteq V_{k-1}\cup W_{k-1},~|V_{k-1}|+|W_{k-1}|=n-|U|,~V_{j}\cup W_{j}=M_{j},~j=1,2,\ldots,\lfloor\frac{k}{2}\rfloor-2,~|V_{\lfloor\frac{k}{2}\rfloor-1}|+|W_{\lfloor\frac{k}{2}\rfloor-1}|=|M_{\lfloor\frac{k}{2}\rfloor-1}|+|M_{\lfloor\frac{k}{2}\rfloor}|-(n-|U|)$ and $|V_{\lfloor\frac{k}{2}\rfloor}|+|W_{\lfloor\frac{k}{2}\rfloor}|=\cdots=|V_{k-2}|+|W_{k-2}|=|V_{k}|+|W_{k}|=0$.
  \item When $0<n-|U|\leq|M_{\lfloor\frac{k}{2}\rfloor}|$, under the premise that the above two conditions are satisfied, the minimum value of $k|U|=(|V_{1}|+|W_{1}|)+\cdots+(k-1)(|V_{k-1}|+|W_{k-1}|)+k(|V_{k}|+|W_{k}|)$ is $(k-1)(n-|U|)+\sum_{j=1}^{\lfloor\frac{k}{2}\rfloor-1}j\cdot|M_{j}|+\lfloor\frac{k}{2}\rfloor\left[|M_{\lfloor\frac{k}{2}\rfloor}|-(n-|U|)\right]$, where $|V_{k-1}|+|W_{k-1}|=n-|U|,~V_{j}\cup W_{j}=M_{j},~j=1,2,\ldots,\lfloor\frac{k}{2}\rfloor-1,~|V_{\lfloor\frac{k}{2}\rfloor}|+|W_{\lfloor\frac{k}{2}\rfloor}|=|M_{\lfloor\frac{k}{2}\rfloor}|-(n-|U|)$ and $|V_{\lfloor\frac{k}{2}\rfloor+1}|+|W_{\lfloor\frac{k}{2}\rfloor+1}|=\cdots=|V_{k-2}|+|W_{k-2}|=|V_{k}|+|W_{k}|=0$.
\end{itemize}
In a word, according to Eq. (\ref{e3}), when $\sum_{j=l}^{\lfloor\frac{k}{2}\rfloor}|M_{j}|<n-|U|\leq\sum_{j=l-1}^{\lfloor\frac{k}{2}\rfloor}|M_{j}|,~l=1,2,\ldots,\lfloor\frac{k}{2}\rfloor+1$, we have
\begin{eqnarray*}
k|U|=(k-l)(n-|U|)+\sum_{j=1}^{l-2}j\cdot|M_{j}|+(l-1)\left(\sum_{j=l-1}^{\lfloor\frac{k}{2}\rfloor}|M_{j}|\right),\quad{\rm for~any}~k_{1}<k_{2},~\sum_{j=k_{2}}^{k_{1}}|M_{j}|=0,
\end{eqnarray*}
Then we note that for some $l^{*}$, when $\sum_{j=l^{*}}^{\lfloor\frac{k}{2}\rfloor}|M_{j}|<n-|U|\leq\sum_{j=l^{*}-1}^{\lfloor\frac{k}{2}\rfloor}|M_{j}|$,
\begin{itemize}
  \item for any $l<l^{*}$, we have
  \begin{equation*}
  \begin{aligned}
  &\left[(k-l^{*})(n-|U|)+\sum_{j=1}^{l^{*}-2}j\cdot|M_{j}|+(l^{*}-1)\left(\sum_{j=l^{*}-1}^{\lfloor\frac{k}{2}\rfloor}|M_{j}|\right)\right]-\left[(k-l)(n-|U|)+\sum_{j=1}^{l-2}j\cdot|M_{j}|\right.\\
  &\left.+(l-1)\left(\sum_{j=l-1}^{\lfloor\frac{k}{2}\rfloor}|M_{j}|\right)\right]\\
  =&-(l^{*}-l)(n-|U|)+(l^{*}-l)\left(\sum_{j=l^{*}-1}^{\lfloor\frac{k}{2}\rfloor}|M_{j}|\right)+\sum_{j=l-1}^{l^{*}-2}j\cdot|M_{j}|-(l-1)\left(\sum_{j=l-1}^{l^{*}-2}|M_{j}|\right)\\
  \geq&-(l^{*}-l)\left(\sum_{j=l^{*}-1}^{\lfloor\frac{k}{2}\rfloor}|M_{j}|\right)+(l^{*}-l)\left(\sum_{j=l^{*}-1}^{\lfloor\frac{k}{2}\rfloor}|M_{j}|\right)+\sum_{j=l-1}^{l^{*}-2}j\cdot|M_{j}|-(l-1)\left(\sum_{j=l-1}^{l^{*}-2}|M_{j}|\right)\\
  \geq&0.
  \end{aligned}
  \end{equation*}
  \item for any $l>l^{*}$, we have
  \begin{equation*}
  \begin{aligned}
  &\left[(k-l^{*})(n-|U|)+\sum_{j=1}^{l^{*}-2}j\cdot|M_{j}|+(l^{*}-1)\left(\sum_{j=l^{*}-1}^{\lfloor\frac{k}{2}\rfloor}|M_{j}|\right)\right]-\left[(k-l)(n-|U|)+\sum_{j=1}^{l-2}j\cdot|M_{j}|\right.\\
  &\left.+(l-1)\left(\sum_{j=l-1}^{\lfloor\frac{k}{2}\rfloor}|M_{j}|\right)\right]\\
  =&(l-l^{*})(n-|U|)-(l-l^{*})\left(\sum_{j=l-1}^{\lfloor\frac{k}{2}\rfloor}|M_{j}|\right)-\sum_{j=l^{*}-1}^{l-2}j\cdot|M_{j}|+(l^{*}-1)\left(\sum_{j=l^{*}-1}^{l-2}|M_{j}|\right)\\
  >&(l-l^{*})\left(\sum_{j=l^{*}}^{\lfloor\frac{k}{2}\rfloor}|M_{j}|\right)-(l-l^{*})\left(\sum_{j=l-1}^{\lfloor\frac{k}{2}\rfloor}|M_{j}|\right)-\sum_{j=l^{*}-1}^{l-2}j\cdot|M_{j}|+(l^{*}-1)\left(\sum_{j=l^{*}-1}^{l-2}|M_{j}|\right)\\
  >&(l-l^{*})\left(\sum_{j=l^{*}}^{l-2}|M_{j}|\right)+(l^{*}-1)\left(\sum_{j=l^{*}-1}^{l-2}|M_{j}|\right)-\sum_{j=l^{*}-1}^{l-2}j\cdot|M_{j}|\\
  =&(l^{*}-1)|M_{l^{*}-1}|+(l-1)\left(\sum_{j=l^{*}}^{l-2}|M_{j}|\right)-\sum_{j=l^{*}-1}^{l-2}j\cdot|M_{j}|\\
  =&(l-1)\left(\sum_{j=l^{*}}^{l-2}|M_{j}|\right)-\sum_{j=l^{*}}^{l-2}j\cdot|M_{j}|\\
  >&0,
  \end{aligned}
  \end{equation*}
\end{itemize}
which means that
\begin{equation*}
k|U|=\max\left\{(k-l)(n-|U|)+\sum_{j=1}^{l-2}j\cdot|M_{j}|+(l-1)\left(\sum_{j=l-1}^{\lfloor\frac{k}{2}\rfloor}|M_{j}|\right),~l=1,\ldots,\lfloor\frac{k}{2}\rfloor+1\right\}.
\end{equation*}

Suppose that when $l=l^{*}$, $(k-l)(n-|U|)+\sum_{j=1}^{l-2}j\cdot|M_{j}|+(l-1)\left(\sum_{j=l-1}^{\lfloor\frac{k}{2}\rfloor}|M_{j}|\right)$ take the maximum value. Then
\begin{eqnarray*}
k|U|&=&\max\left\{(k-l)(n-|U|)+\sum_{j=1}^{l-2}j\cdot|M_{j}|+(l-1)\left(\sum_{j=l-1}^{\lfloor\frac{k}{2}\rfloor}|M_{j}|\right),~l=1,\ldots,\lfloor\frac{k}{2}\rfloor+1\right\}\\
&=&(k-l^{*})(n-|U|)+\sum_{j=1}^{l^{*}-2}j\cdot|M_{j}|+(l^{*}-1)\left(\sum_{j=l^{*}-1}^{\lfloor\frac{k}{2}\rfloor}|M_{j}|\right),
\end{eqnarray*}
and thus, we have
\begin{equation*}
|U|=\frac{k-l^{*}}{2k-l^{*}}n+\frac{1}{2k-l^{*}}\left(\sum_{j=1}^{l^{*}-2}j\cdot|M_{j}|\right)+\frac{l^{*}-1}{2k-l^{*}}\left(\sum_{j=l^{*}-1}^{\lfloor\frac{k}{2}\rfloor}|M_{j}|\right).
\end{equation*}
Now, for any $l\neq l^{*}$
\begin{footnotesize}
\tiny
\begin{equation*}
\begin{aligned}
(k-l^{*})(n-|U|)+\sum_{j=1}^{l^{*}-2}j\cdot|M_{j}|+(l^{*}-1)\left(\sum_{j=l^{*}-1}^{\lfloor\frac{k}{2}\rfloor}|M_{j}|\right) &\geq (k-l)(n-|U|)+\sum_{j=1}^{l-2}j\cdot|M_{j}|+(l-1)\left(\sum_{j=l-1}^{\lfloor\frac{k}{2}\rfloor}|M_{j}|\right)\\
(l^{*}-1)\left(\sum_{j=l^{*}-1}^{\lfloor\frac{k}{2}\rfloor}|M_{j}|\right)-(k-l^{*})|U| &\geq (k-l)n+\sum_{j=1}^{l-2}j\cdot|M_{j}|+(l-1)\left(\sum_{j=l-1}^{\lfloor\frac{k}{2}\rfloor}|M_{j}|\right)-(k-l)|U|\\
(k-l^{*})n+\sum_{j=1}^{l^{*}-2}j\cdot|M_{j}|+(l^{*}-1)\left(\sum_{j=l^{*}-1}^{\lfloor\frac{k}{2}\rfloor}|M_{j}|\right)-(2k-l^{*})|U| &\geq (k-l)n+\sum_{j=1}^{l-2}j\cdot|M_{j}|+(l-1)\left(\sum_{j=l-1}^{\lfloor\frac{k}{2}\rfloor}|M_{j}|\right)-(2k-l)|U|,
\end{aligned}
\end{equation*}
\end{footnotesize}
and then we have $(2k-l^{*})\left[\frac{k-l^{*}}{2k-l^{*}}n+\frac{1}{2k-l^{*}}\left(\sum_{j=1}^{l^{*}-2}j\cdot|M_{j}|\right)+\frac{l^{*}-1}{2k-l^{*}}\left(\sum_{j=l^{*}-1}^{\lfloor\frac{k}{2}\rfloor}|M_{j}|\right)-|U|\right] \geq (2k-l)\left[\frac{k-l}{2k-l}n+\frac{1}{2k-l}\left(\sum_{j=1}^{l-2}j\cdot|M_{j}|\right)+\frac{l-1}{2k-l}\left(\sum_{j=l-1}^{\lfloor\frac{k}{2}\rfloor}|M_{j}|\right)-|U|\right]$, which means that
\begin{equation*}
\begin{aligned}
0 &\geq (2k-l)\left[\frac{k-l}{2k-l}n+\frac{1}{2k-l}\left(\sum_{j=1}^{l-2}j\cdot|M_{j}|\right)+\frac{l-1}{2k-l}\left(\sum_{j=l-1}^{\lfloor\frac{k}{2}\rfloor}|M_{j}|\right)-|U|\right]\\
|U| &\geq   \frac{k-l}{2k-l}n+\frac{1}{2k-l}\left(\sum_{j=1}^{l-2}j\cdot|M_{j}|\right)+\frac{l-1}{2k-l}\left(\sum_{j=l-1}^{\lfloor\frac{k}{2}\rfloor}|M_{j}|\right),
\end{aligned}
\end{equation*}
so,
\begin{eqnarray*}
|U|=\max\left\{\frac{k-l}{2k-l}n+\frac{1}{2k-l}\left(\sum_{j=1}^{l-2}j\cdot|M_{j}|\right)+\frac{l-1}{2k-l}\left(\sum_{j=l-1}^{\lfloor\frac{k}{2}\rfloor}|M_{j}|\right),~l=1,2,\ldots,\lfloor\frac{k}{2}\rfloor+1\right\}.
\end{eqnarray*}

Therefore, in order to satisfy the controllability condition, the number of nodes we need to control is at least
\begin{eqnarray*}
\max\left\{\frac{k-l}{2k-l}n+\frac{1}{2k-l}\left(\sum_{j=1}^{l-2}j\cdot|M_{j}|\right)+\frac{l-1}{2k-l}\left(\sum_{j=l-1}^{\lfloor\frac{k}{2}\rfloor}|M_{j}|\right),~l=1,2,\ldots,\lfloor\frac{k}{2}\rfloor+1\right\}.
\end{eqnarray*}
\qed

\begin{remark}
It is noted that Theorem \ref{thm:2-2-AND-negation} is a special case of Theorem \ref{thm:k-k-AND-negation}, that is, by letting $k=2$ and $l=1,2$, the general lower bound on the size of Theorem \ref{thm:k-k-AND-negation} becomes $\max\left(\frac{1}{3}n,\frac{1}{2}|M_{1}|\right)$. On the other hand, Theorem \ref{thm:k-k-AND} is also a special case of Theorem \ref{thm:k-k-AND-negation}, that is, by letting $|M_{0}|=n$, the general lower bound on the size of Theorem \ref{thm:k-k-AND-negation} becomes  $\frac{k-1}{2k-1}n$ (i.e., $\max\left\{\frac{k-l}{2k-l}n,~l=1,\ldots,\lfloor\frac{k}{2}\rfloor+1\right\}=\frac{k-1}{2k-1}n$).
Additionally, when $|M_{0}|<n$, the general lower bound given in Theorem \ref{thm:k-k-AND-negation} is larger than the general lower bound given in Theorem \ref{thm:k-k-AND}.
\end{remark}

\begin{example}\label{example6}
Construct a 4-4-AND-BN with negation as follows:
\begin{eqnarray*}
x_1(t+1) & = & \lnon{x_5(t)} \land x_6(t) \land x_{7}(t) \land \lnon{x_{8}(t)},\\
x_2(t+1) & = & \lnon{x_5(t)} \land x_6(t) \land x_{7}(t) \land \lnon{x_{8}(t)},\\
x_3(t+1) & = & \lnon{x_5(t)} \land x_6(t) \land x_{7}(t) \land \lnon{x_{8}(t)},\\
x_4(t+1) & = & x_1(t) \land \lnon{x_2(t)} \land x_{3}(t) \land x_{12}(t),\\
x_5(t+1) & = & \lnon{x_1(t)} \land x_2(t) \land \lnon{x_3(t)} \land \lnon{x_4(t)},\\
x_6(t+1) & = & \lnon{x_2(t)} \land x_3(t) \land x_4(t) \land x_5(t),\\
x_7(t+1) & = & x_1(t) \land x_3(t) \land x_4(t) \land \lnon{x_6(t)},\\
x_8(t+1) & = & x_1(t) \land \lnon{x_2(t)} \land x_4(t) \land \lnon{x_7(t)},\\
x_9(t+1) & = & x_8(t) \land x_{13}(t) \land x_{14}(t) \land x_{15}(t),\\
x_{10}(t+1) & = & \lnon{x_9(t)} \land x_{13}(t) \land x_{14}(t) \land x_{15}(t),\\
x_{11}(t+1) & = & \lnon{x_{10}(t)} \land x_{13}(t) \land x_{14}(t) \land x_{15}(t),\\
x_{12}(t+1) & = & x_{11}(t) \land \lnon{x_{13}(t)} \land \lnon{x_{14}(t)} \land \lnon{x_{15}(t)},\\
x_{13}(t+1) & = & x_{9}(t) \land x_{10}(t) \land \lnon{x_{11}(t)} \land \lnon{x_{12}(t)},\\
x_{14}(t+1) & = & x_{9}(t) \land x_{10}(t) \land \lnon{x_{11}(t)} \land \lnon{x_{12}(t)},\\
x_{15}(t+1) & = & x_{9}(t) \land x_{10}(t) \land \lnon{x_{11}(t)} \land \lnon{x_{12}(t)}
\end{eqnarray*}
with letting $x_{1},x_{2},x_{3},x_{5},x_{12},x_{13},x_{14},x_{15}$ be control nodes. By using control signals, we let $x_{1}(t)=x_{3}(t)=x_{13}(t)=x_{14}(t)=x_{15}(t)=1$ and $x_{2}(t)=0$ for all $t=1,2,\ldots,6$, $x_{12}(1)=x_{12}(2)=\cdots=x_{12}(5)=1,~x_{12}(6)=x_{4}^{T}$, and $x_{1}(7)=x_{1}^{T},x_{2}(7)=x_{2}^{T},x_{3}(7)=x_{3}^{T},x_{5}(7)=x_{5}^{T},x_{12}(7)=x_{12}^{T},x_{13}(7)=x_{13}^{T},x_{14}(7)=x_{14}^{T},x_{15}(7)=x_{15}^{T}$. Let $x_{5}(1)=x_{11}^{T},~\lnon{x_{5}(2)}=x_{10}^{T},~x_{5}(3)=x_{9}^{T},~x_{5}(4)=x_{8}^{T},~\lnon{x_{5}(5)}=x_{7}^{T}$ and $x_{5}(6)=x_{6}^{T}$. Then, it is straightforward to see that $\xvec(7)=\xvec^T$.

In this case, for any $x_{i}\in\{x_{1},x_{3},x_{4},x_{6},x_{7},x_{9},x_{10},x_{13},x_{14},x_{15}\}$, the literal $\lnon{x_{i}}$ appears 1 time in the 4-4-AND-BN with negation; for any $x_{i}\in\{x_{2},x_{5},x_{8},x_{11},x_{12}\}$, the literal $x_{i}$ appears 1 time in the 4-4-AND-BN with negation, so $M_{1}=V$ ($|M_{1}|=15$).

It can be seen that $U=\{x_{1},x_{2},x_{3},x_{5},x_{12},x_{13},x_{14},x_{15}\}$ ($|U|=8$) is the minimum control node set. Suppose there is a control node set $U'$, where $|U'|=7$, then according to the conditions (I) and (II) in the proof of Theorem \ref{thm:k-k-AND-negation}, we need to satisfy $|V_{3}|+|W_{3}|\geq 8$ and for any node $x_{i}\in M_{1}$, $|\Gamma^{+}(x_{i})\cap U|\geq 1$. Hence, the value of $(|V_{1}|+|W_{1}|)+2(|V_{2}|+|W_{2}|)+3(|V_{3}|+|W_{3}|)$ is not less than $(15-8)+3\times8=31$. However, $4|U'|=28<31$, which contradicts Eq. (\ref{e3}), so $U=\{x_{1},x_{2},x_{3},x_{5},x_{12},x_{13},x_{14},x_{15}\}$ is the minimum control node set.

On the other hand, based on Theorem \ref{thm:k-k-AND-negation}, we deduce that the number of control nodes is at least
\begin{eqnarray*}
\max\left\{\frac{3}{7}n,\frac{1}{3}n+\frac{1}{6}(|M_{1}|+|M_{2}|),\frac{1}{5}n+\frac{1}{5}|M_{1}|+\frac{2}{5}|M_{2}|\right\},\quad{\rm where}~n=15,|M_{1}|=15,|M_{2}|=0,
\end{eqnarray*}
that is, $\max(\frac{45}{7},\frac{15}{2},6)=\frac{15}{2}$, which is consistent with the fact that $U=\{x_{1},x_{2},x_{3},x_{5},x_{12},x_{13},x_{14},x_{15}\}$ ($|U|=8$) is the minimum control node set.
\end{example}

\begin{example}\label{example7}
Construct a 4-4-AND-BN with negation as follows:
\begin{eqnarray*}
x_1(t+1) & = & \lnon{x_5(t)} \land x_6(t) \land \lnon{x_{7}(t)} \land x_{8}(t),\\
x_2(t+1) & = & \lnon{x_5(t)} \land \lnon{x_6(t)} \land x_{7}(t) \land \lnon{x_{8}(t)},\\
x_3(t+1) & = & x_5(t) \land \lnon{x_6(t)} \land \lnon{x_{7}(t)} \land \lnon{x_{8}(t)},\\
x_4(t+1) & = & \lnon{x_1(t)} \land x_2(t) \land \lnon{x_{3}(t)} \land x_{4}(t),\\
x_5(t+1) & = & \lnon{x_1(t)} \land x_2(t) \land \lnon{x_3(t)} \land x_4(t),\\
x_6(t+1) & = & \lnon{x_2(t)} \land x_3(t) \land x_5(t) \land x_{12}(t),\\
x_7(t+1) & = & x_1(t) \land x_3(t) \land \lnon{x_4(t)} \land x_6(t),\\
x_8(t+1) & = & x_1(t) \land \lnon{x_2(t)} \land \lnon{x_4(t)} \land x_7(t),\\
x_9(t+1) & = & x_8(t) \land x_{13}(t) \land \lnon{x_{14}(t)} \land x_{15}(t),\\
x_{10}(t+1) & = & x_9(t) \land x_{13}(t) \land \lnon{x_{14}(t)} \land x_{15}(t),\\
x_{11}(t+1) & = & x_{10}(t) \land x_{13}(t) \land \lnon{x_{14}(t)} \land x_{15}(t),\\
x_{12}(t+1) & = & x_{11}(t) \land \lnon{x_{13}(t)} \land x_{14}(t) \land \lnon{x_{15}(t)},\\
x_{13}(t+1) & = & \lnon{x_{9}(t)} \land x_{10}(t) \land \lnon{x_{11}(t)} \land \lnon{x_{12}(t)},\\
x_{14}(t+1) & = & \lnon{x_{9}(t)} \land \lnon{x_{10}(t)} \land x_{11}(t) \land x_{12}(t),\\
x_{15}(t+1) & = & x_{9}(t) \land \lnon{x_{10}(t)} \land \lnon{x_{11}(t)} \land \lnon{x_{12}(t)}
\end{eqnarray*}
with letting $x_{1},x_{2},x_{3},x_{4},x_{5},x_{12},x_{13},x_{14},x_{15}$ be control nodes. By using control signals, we let $x_{1}(t)=x_{3}(t)=x_{5}(t)=x_{13}(t)=x_{15}(t)=1$ and $x_{2}(t)=x_{4}(t)=x_{14}(t)=0$ for all $t=1,2,\ldots,6$, and $x_{1}(7)=x_{1}^{T},x_{2}(7)=x_{2}^{T},x_{3}(7)=x_{3}^{T},x_{4}(7)=x_{4}^{T},x_{5}(7)=x_{5}^{T},x_{12}(7)=x_{12}^{T},x_{13}(7)=x_{13}^{T},x_{14}(7)=x_{14}^{T},x_{15}(7)=x_{15}^{T}$. Let $x_{12}(t)=x_{12-t}^{T},~t=1,2,\ldots,6$. Then, it is straightforward to see that $\xvec(7)=\xvec^T$.

In this case, for any $x_{i}\in\{x_{1},x_{2},x_{3},x_{4},x_{5},x_{6},x_{7},x_{8},x_{9},x_{10},x_{11},x_{12}\}$, the literal $\lnon{x_{i}}$ (resp., $x_{i}$) appears 2 times (resp., 2 times) in the 4-4-AND-BN with negation; for any $x_{i}\in\{x_{13},x_{15}\}$ (resp., $x_{i}\in\{x_{14}\}$), the literal $\lnon{x_{i}}$ (resp., $x_{14}$) appears 1 time in the 4-4-AND-BN with negation, so $M_{1}=\{x_{13},x_{14},x_{15}\}$ and $M_{2}=\{x_{1},x_{2},x_{3},x_{4},x_{5},x_{6},x_{7},x_{8},x_{9},x_{10},x_{11},x_{12}\}$ ($|M_{1}|=3$ and $|M_{2}|=12$).

It can be seen that $U=\{x_{1},x_{2},x_{3},x_{4},x_{5},x_{12},x_{13},x_{14},x_{15}\}$ ($|U|=9$) is the minimum control node set. Suppose there is a control node set $U'$, where $|U'|=8$, then according to the conditions (I) and (II) in the proof of Theorem \ref{thm:k-k-AND-negation}, we need to satisfy $|V_{3}|+|W_{3}|\geq 7$ and for any node $x_{i}\in M_{1}$ (resp., $x_{i}\in M_{2}$), $|\Gamma^{+}(x_{i})\cap U|\geq 1$ (reps. $|\Gamma^{+}(x_{i})\cap U|\geq 2$). Hence, the value of $(|V_{1}|+|W_{1}|)+2(|V_{2}|+|W_{2}|)+3(|V_{3}|+|W_{3}|)$ is not less than $3+2\times(12-7)+3\times7=34$. However, $4|U'|=32<34$, which contradicts Eq. (\ref{e3}), so $U=\{x_{1},x_{2},x_{3},x_{4},x_{5},x_{12},x_{13},x_{14},x_{15}\}$ is the minimum control node set.

On the other hand, based on Theorem \ref{thm:k-k-AND-negation}, we deduce that the number of control nodes is at least
\begin{eqnarray*}
\max\left\{\frac{3}{7}n,\frac{1}{3}n+\frac{1}{6}(|M_{1}|+|M_{2}|),\frac{1}{5}n+\frac{1}{5}|M_{1}|+\frac{2}{5}|M_{2}|\right\},\quad{\rm where}~n=15,|M_{1}|=3,|M_{2}|=12,
\end{eqnarray*}
that is, $\max(\frac{45}{7},\frac{15}{2},\frac{42}{5})=\frac{42}{5}$, which is consistent with the fact that
\begin{eqnarray*}
U=\{x_{1},x_{2},x_{3},x_{4},x_{5},x_{12},x_{13},x_{14},x_{15}\}~(|U|=9)
\end{eqnarray*}
is the minimum control node set.
\end{example}

\begin{remark}\label{remark10}
According to Examples \ref{example6} and \ref{example7}, in the case of $n=15$ and $k=4$, it can be seen that $\frac{1}{3}n+\frac{1}{6}(|M_{1}|+|M_{2}|)>\frac{3}{7}n$ when $|M_{1}|+|M_{2}|>8$, and $\frac{1}{5}n+\frac{1}{5}|M_{1}|+\frac{2}{5}|M_{2}|>\frac{1}{3}n+\frac{1}{6}(|M_{1}|+|M_{2}|)$ when $|M_{1}|+7|M_{2}|>60$, which means that the larger $|M_{1}|$ and $|M_{2}|$, the larger the general lower bound on the size of control node set, and $|M_{2}|$ has a greater impact on the value of the general lower bound.

More generally, according to Theorem \ref{thm:k-k-AND-negation}, the number of control nodes is at least
\begin{equation*}
\begin{aligned}
\max\left\{\frac{k-l}{2k-l}n+\frac{1}{2k-l}\left(\sum_{j=1}^{l-2}j\cdot|M_{j}|\right)+\frac{l-1}{2k-l}\left(\sum_{j=l-1}^{\lfloor\frac{k}{2}\rfloor}|M_{j}|\right),~l=1,\ldots,\lfloor\frac{k}{2}\rfloor+1\right\},
\end{aligned}
\end{equation*}
and then for each $l=l^{*}$, it follows that
\begin{eqnarray*}
&&\left\{\frac{k-(l^{*}+1)}{2k-(l^{*}+1)}n+\frac{1}{2k-(l^{*}+1)}\left(\sum_{j=1}^{(l^{*}+1)-2}j\cdot|M_{j}|\right)+\frac{(l^{*}+1)-1}{2k-(l^{*}+1)}\left(\sum_{j=(l^{*}+1)-1}^{\lfloor\frac{k}{2}\rfloor}|M_{j}|\right)\right\}-\\
&&\left\{\frac{k-l^{*}}{2k-l^{*}}n+\frac{1}{2k-l^{*}}\left(\sum_{j=1}^{l^{*}-2}j\cdot|M_{j}|\right)+\frac{l^{*}-1}{2k-l^{*}}\left(\sum_{j=l^{*}-1}^{\lfloor\frac{k}{2}\rfloor}|M_{j}|\right)\right\}\\
&=&\frac{-kn}{[2k-(l^{*}+1)](2k-l^{*})}+\frac{(2k-l^{*})(l^{*}-1)|M_{l^{*}-1}|+\sum_{j=1}^{l^{*}-2}j\cdot|M_{j}|}{[2k-(l^{*}+1)](2k-l^{*})}+\\
&&\frac{-(2k-l^{*}-1)(l^{*}-1)|M_{l^{*}-1}|+(2k-1)\sum_{j=l^{*}}^{\lfloor\frac{k}{2}\rfloor}|M_{j}|}{[2k-(l^{*}+1)](2k-l^{*})}\\
%&=&\frac{-kn}{[2k-(l^{*}+1)](2k-l^{*})}+\frac{(l^{*}-1)|M_{l^{*}-1}|+\sum_{j=1}^{l^{*}-2}j\cdot|M_{j}|+(2k-1)\sum_{j=l^{*}}^{\lfloor\frac{k}{2}\rfloor}|M_{j}|}{[2k-(l^{*}+1)](2k-l^{*})}\\
&=&\frac{-kn}{[2k-(l^{*}+1)](2k-l^{*})}+\frac{\sum_{j=1}^{l^{*}-1}j\cdot|M_{j}|+(2k-1)\sum_{j=l^{*}}^{\lfloor\frac{k}{2}\rfloor}|M_{j}|}{[2k-(l^{*}+1)](2k-l^{*})},
\end{eqnarray*}
which implies that $\frac{k-(l^{*}+1)}{2k-(l^{*}+1)}n+\frac{1}{2k-(l^{*}+1)}\left(\sum_{j=1}^{(l^{*}+1)-2}j\cdot|M_{j}|\right)+\frac{(l^{*}+1)-1}{2k-(l^{*}+1)}\left(\sum_{j=(l^{*}+1)-1}^{\lfloor\frac{k}{2}\rfloor}|M_{j}|\right)>\frac{k-l^{*}}{2k-l^{*}}n+\frac{1}{2k-l^{*}}\left(\sum_{j=1}^{l^{*}-2}j\cdot|M_{j}|\right)+\frac{l^{*}-1}{2k-l^{*}}\left(\sum_{j=l^{*}-1}^{\lfloor\frac{k}{2}\rfloor}|M_{j}|\right)$,
when $\sum_{j=1}^{l^{*}-1}j\cdot|M_{j}|+(2k-1)(|M_{l^{*}}|+\cdots+|M_{\lfloor\frac{k}{2}\rfloor}|)>kn$. Hence, we can deduce that
\begin{itemize}
  \item $\frac{k-2}{2k-2}n+\frac{1}{2k-2}\left(\sum_{j=1}^{\lfloor\frac{k}{2}\rfloor}|M_{j}|\right)>\frac{k-1}{2k-1}n$,
when $(2k-1)(|M_{1}|+\cdots+|M_{\lfloor\frac{k}{2}\rfloor}|)>kn$,
  \item $\frac{k-3}{2k-3}n+\frac{1}{2k-3}|M_{1}|+\frac{2}{2k-3}\left(\sum_{j=2}^{\lfloor\frac{k}{2}\rfloor}|M_{j}|\right)>\frac{k-2}{2k-2}n+\frac{1}{2k-2}\left(\sum_{j=1}^{\lfloor\frac{k}{2}\rfloor}|M_{j}|\right)$,
when $|M_{1}|+(2k-1)(|M_{2}|+\cdots+|M_{\lfloor\frac{k}{2}\rfloor}|)>kn$,
  \item $\ldots$
  \item $\frac{k-(\lfloor\frac{k}{2}\rfloor+1)}{2k-(\lfloor\frac{k}{2}\rfloor+1)}n+\frac{1}{2k-(\lfloor\frac{k}{2}\rfloor+1)}\left(\sum_{j=1}^{(\lfloor\frac{k}{2}\rfloor+1)-2}j\cdot|M_{j}|\right)+\frac{(\lfloor\frac{k}{2}\rfloor+1)-1}{2k-(\lfloor\frac{k}{2}\rfloor+1)}\left(\sum_{j=(\lfloor\frac{k}{2}\rfloor+1)-1}^{\lfloor\frac{k}{2}\rfloor}|M_{j}|\right)>\frac{k-\lfloor\frac{k}{2}\rfloor}{2k-\lfloor\frac{k}{2}\rfloor}n+\frac{1}{2k-\lfloor\frac{k}{2}\rfloor}\left(\sum_{j=1}^{\lfloor\frac{k}{2}\rfloor-2}j\cdot|M_{j}|\right)+\frac{\lfloor\frac{k}{2}\rfloor-1}{2k-\lfloor\frac{k}{2}\rfloor}\left(\sum_{j=\lfloor\frac{k}{2}\rfloor-1}^{\lfloor\frac{k}{2}\rfloor}|M_{j}|\right)$,
when $\sum_{j=1}^{\lfloor\frac{k}{2}\rfloor-1}j\cdot|M_{j}|+(2k-1)|M_{\lfloor\frac{k}{2}\rfloor}|>kn$,
\end{itemize}
which means that the larger $|M_{j}|,~j=1,\ldots,\lfloor\frac{k}{2}\rfloor$, the larger the general lower bound on the size of control node set, and the influence of the values of $|M_{1}|,\ldots,|M_{\lfloor\frac{k}{2}\rfloor}|$ on the value of the general lower bound is from small to large, that is, $|M_{\lfloor\frac{k}{2}\rfloor}|$ has the greatest impact on the value of the general lower bound.
\end{remark}

\begin{remark}
The above result for $k$-$k$-AND-BNs with negation ($k\geq2$) also applies to $k$-$k$-OR-BNs with negation ($k\geq2$).
\end{remark}
\section{Boolean networks consisting of nested canalyzing functions}
Finally, we consider the case where the nested canalyzing function is assigned to each node, and we refer to this BN as a $k$-$k$-NC-BN. The idea behind the following proposition is to use a kind of shift register
again, where the details are substantially different from the previous ones.

\begin{proposition}\label{prop:nc}
For any integer $n>2$ and $(n \mod k)=0$, there exists a $k$-$k$-NC-BN for which the size of the minimum control node set is $\frac{n}{k}$.
\end{proposition}
(Proof)
We construct a $k$-$k$-NC-BN, where $n=k$ by
\begin{eqnarray*}
x_1(t+1) & = & x_2(t) \lor (\lnon{x_1(t)} \land \lnon{x_3(t)} \land \lnon{x_4(t)} \land \cdots \land \lnon{x_n(t)}),\\
x_2(t+1) & = & x_3(t) \lor (\lnon{x_1(t)} \land \lnon{x_2(t)} \land \lnon{x_4(t)} \land \cdots \land \lnon{x_n(t)}),\\
&\vdots & \\
x_{n-1}(t+1) & = & x_{n}(t) \lor (\lnon{x_1(t)} \land \lnon{x_2(t)} \land \cdots \land \lnon{x_{n-2}(t)} \land \lnon{x_{n-1}(t)}),\\
x_{n}(t+1) & = & x_{1}(t) \lor (\lnon{x_2(t)} \land \lnon{x_3(t)} \land \lnon{x_4(t)} \land \cdots \land \lnon{x_n(t)})
\end{eqnarray*}
with letting $x_{1}$ be the control node. By using control signals, we let $x_{1}(1)=1$, $x_{1}(t)=x_{t}^{T}$ for $t=2,\ldots,n$ and $x_{1}(n+1)=x_{1}^{T}$.

Note that when $x_{1}(1)=1$, we have $x_{n}(2)=1$, and then we get the following results,
\begin{itemize}
    \item When $t=3$, we have $x_{2}(3)=x_{3}(2),~x_{3}(3)=x_{4}(2),\ldots,x_{n-2}(3)=x_{n-1}(2),~x_{n-1}(3)=x_{n}(2)=1,~x_{n}(3)=x_{1}(2)=x_{2}^{T}$;
     \item When $t=4$, we have $x_{2}(4)=x_{3}(3),~x_{3}(4)=x_{4}(3),\ldots,x_{n-3}(4)=x_{n-2}(3),~x_{n-2}(4)=x_{n-1}(3)=1,~x_{n-1}(4)=x_{n}(3)=x_{2}^{T},~x_{n}(4)=x_{1}(3)=x_{3}^{T}$;
     \item When $t=5$, we have $x_{2}(5)=x_{3}(4),~x_{3}(5)=x_{4}(4),\ldots,x_{n-4}(5)=x_{n-3}(4),~x_{n-3}(5)=x_{n-2}(4)=1,~
        x_{n-2}(5)=x_{n-1}(4)=x_{2}^{T},~x_{n-1}(5)=x_{n}(4)=x_{3}^{T},~x_{n}(5)=x_{1}(4)=x_{4}^{T}$;
     \item $\ldots$
     \item When $t=n$, we have $x_{2}(n)=1,~x_{3}(n)=x_{2}^{T},~x_{4}(n)=x_{3}^{T},\ldots,x_{n-1}(n)=x_{n-2}^{T},~x_{n}(n)=x_{n-1}^{T}$;
     \item When $t=n+1$, we have $x_{2}(n+1)=x_{3}(n)=x_{2}^{T},~x_{3}(n+1)=x_{4}(n)=x_{3}^{T},~x_{4}(n+1)=x_{5}(n)=x_{4}^{T},\ldots,x_{n-1}(n+1)=x_{n}(n)=x_{n-1}^{T},~x_{n}(n+1)=x_{1}(n)=x_{n}^{T}$.
\end{itemize}
Then, it is straightforward to see that $\xvec(n+1)=\xvec^T$.

When $(n \mod k)=0$, we partition $[x_1,x_2,\ldots,x_n]$
into $\frac{n}{k}$ blocks:
$$[x_1,x_2,\ldots,x_k],[x_{k+1},x_{k+2},\ldots,x_{2k}],\ldots,$$
construct a partial BN for each block as in the first case, then merge these BNs, and let
$$
U=\{x_{1},x_{k+1},\ldots,x_{n-k+1}\}
$$
Clearly, this $k$-$k$-NC-BN is controllable, where $(n \mod k)=0$.
\qed

\begin{remark}
Notably, the minimum size control node set $\frac{n}{k}$ in Proposition \ref{prop:nc} is less than $\frac{k-1}{2k-1}n$ in Theorem \ref{thm:k-k-AND}.
\end{remark}

\begin{remark}
On the other hand, there exists a $k$-$k$-NC-BN whose minimum control node set is larger than that given by Theorem \ref{thm:k-k-AND}. For example, we construct a 3-3-NC-BN as follows:
\begin{eqnarray*}
x_1(t+1) = x_1(t) \lor (\lnon{x_2(t)} \land \lnon{x_3(t)}),\\
x_2(t+1) = x_2(t) \lor (\lnon{x_3(t)} \land \lnon{x_4(t)}),\\
x_3(t+1) = x_3(t) \lor (\lnon{x_4(t)} \land \lnon{x_5(t)}),\\
x_4(t+1) = x_4(t) \lor (\lnon{x_5(t)} \land \lnon{x_1(t)}),\\
x_5(t+1) = x_5(t) \lor (\lnon{x_1(t)} \land \lnon{x_2(t)}).
\end{eqnarray*}
Suppose that $\xvec^0=[1,1,1,1,1]$ and
$\xvec^T=[0,0,0,0,0]$.
It is straightforward to see that the size
of the minimum control node set must be $5$
(i.e., all nodes are control nodes
because $x_i(t)=1$ would hold for all $t$
regardless of the state of other nodes
unless $x_i$ is a control node),
which is larger than that ($\frac{k-1}{2k-1}n=2$) given by Theorem \ref{thm:k-k-AND}.
\end{remark}

\begin{example}
According to Proposition \ref{prop:nc}, we can construct a 5-5-NC-BN, where $n=10$ as follows:
\begin{eqnarray*}
x_1(t+1) & = & x_2(t) \lor (\lnon{x_1(t)} \land \lnon{x_3(t)} \land \lnon{x_4(t)} \land \lnon{x_5(t)}),\\
x_2(t+1) & = & x_3(t) \lor (\lnon{x_1(t)} \land \lnon{x_2(t)} \land \lnon{x_4(t)} \land \lnon{x_5(t)}),\\
x_3(t+1) & = & x_4(t) \lor (\lnon{x_1(t)} \land \lnon{x_2(t)} \land \lnon{x_3(t)} \land \lnon{x_5(t)}),\\
x_4(t+1) & = & x_5(t) \lor (\lnon{x_1(t)} \land \lnon{x_2(t)} \land \lnon{x_3(t)} \land \lnon{x_4(t)}),\\
x_5(t+1) & = & x_1(t) \lor (\lnon{x_2(t)} \land \lnon{x_3(t)} \land \lnon{x_4(t)} \land \lnon{x_5(t)}),\\
x_6(t+1) & = & x_7(t) \lor (\lnon{x_6(t)} \land \lnon{x_8(t)} \land \lnon{x_9(t)} \land \lnon{x_{10}(t)}),\\
x_7(t+1) & = & x_8(t) \lor (\lnon{x_6(t)} \land \lnon{x_7(t)} \land \lnon{x_9(t)} \land \lnon{x_{10}(t)}),\\
x_8(t+1) & = & x_9(t) \lor (\lnon{x_6(t)} \land \lnon{x_7(t)} \land \lnon{x_8(t)} \land \lnon{x_{10}(t)}),\\
x_9(t+1) & = & x_{10}(t) \lor (\lnon{x_6(t)} \land \lnon{x_7(t)} \land \lnon{x_8(t)} \land \lnon{x_{9}(t)}),\\
x_{10}(t+1) & = & x_6(t) \lor (\lnon{x_7(t)} \land \lnon{x_8(t)} \land \lnon{x_9(t)} \land \lnon{x_{10}(t)}).
\end{eqnarray*}
In this case, let $U=\{x_{1},x_{6}\}$. By using control signals, we let $x_{1}(1)=1$, $x_{1}(t)=x_{t}^{T}$ for $t=2,\ldots,5$ and $x_{1}(6)=x_{1}^{T}$. Similarly, let $x_{6}(1)=1$, $x_{6}(t)=x_{5+t}^{T}$ for $t=2,\ldots,5$ and $x_{6}(6)=x_{6}^{T}$. Then, it is straightforward to see that $\xvec(6)=\xvec^T$.

Suppose $\xvec^0=[0,0,0,0,0,0,0,0,0,0]$ and $\xvec^T=[1,0,1,0,1,0,1,0,1,1]$. By using the above control signals, we have the following.
\begin{center}
\begin{tabular}{c|llllllllll}
\hline
$\xvec^0$ ($t=0$) & 0 & 0 & 0 & 0 & 0 & 0 & 0 & 0 & 0 & 0 \\
$\xvec(1)$ & 1 & 1 & 1 & 1 & 1 & 1 & 1 & 1 & 1 & 1 \\
$\xvec(2)$ & 0 & 1 & 1 & 1 & 1 & 1 & 1 & 1 & 1 & 1 \\
$\xvec(3)$ & 1 & 1 & 1 & 1 & 0 & 0 & 1 & 1 & 1 & 1 \\
$\xvec(4)$ & 0 & 1 & 1 & 0 & 1 & 1 & 1 & 1 & 1 & 0 \\
$\xvec(5)$ & 1 & 1 & 0 & 1 & 0 & 1 & 1 & 1 & 0 & 1 \\
$\xvec^T$ ($t=6$) & 1 & 0 & 1 & 0 & 1 & 0 & 1 & 0 & 1 & 1 \\
\hline
\end{tabular}
\end{center}
\end{example}

\section{Simulation results}
\begin{figure}[htbp]
\centering
\subfigure[]{\includegraphics[width=0.45\textwidth]{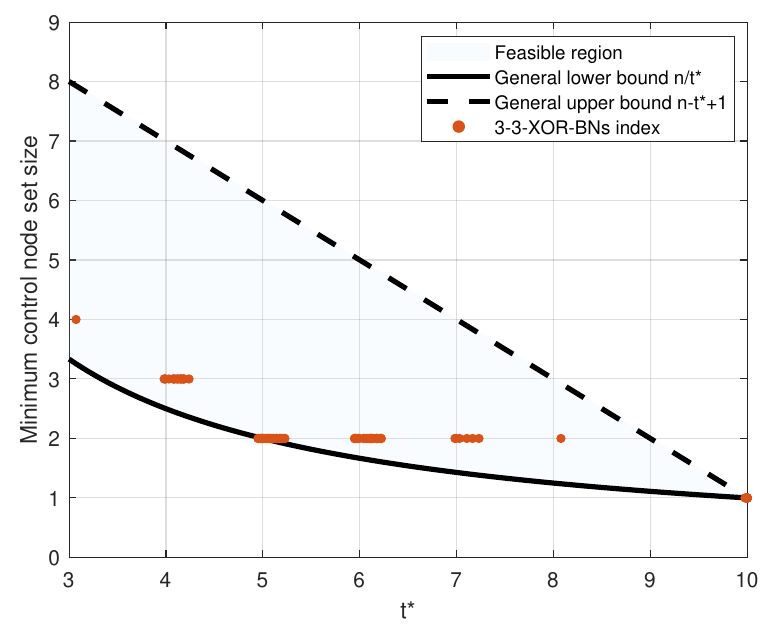}}
\hfil
\subfigure[]{\includegraphics[width=0.45\textwidth]{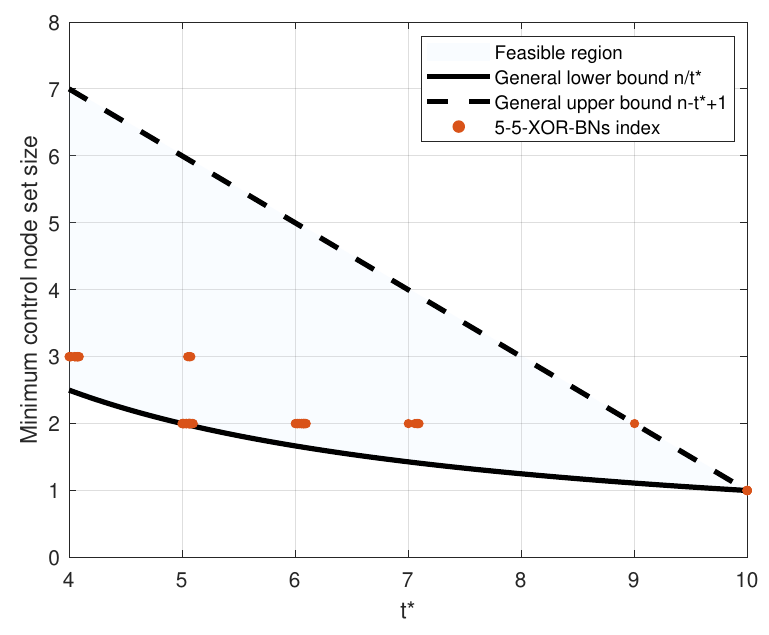}}
\hfil
\subfigure[]{\includegraphics[width=0.45\textwidth]{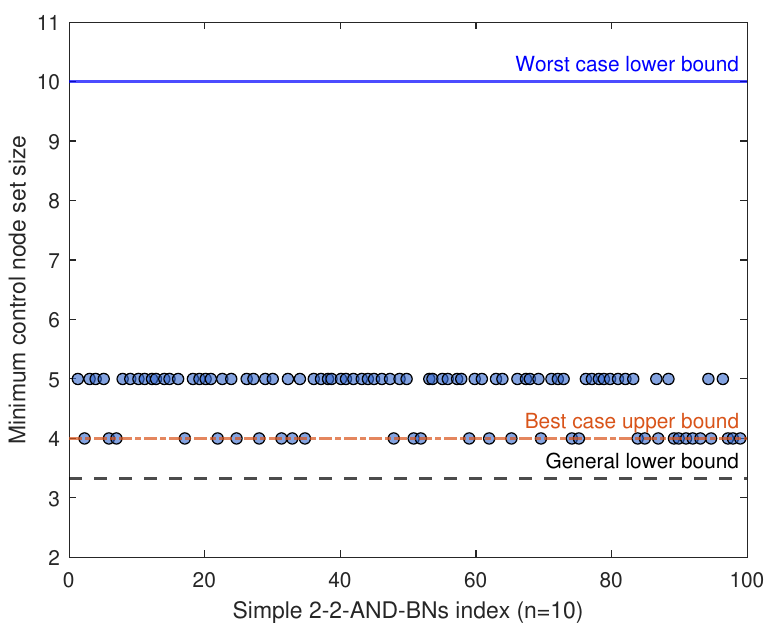}}
\hfil
\subfigure[]{\includegraphics[width=0.45\textwidth]{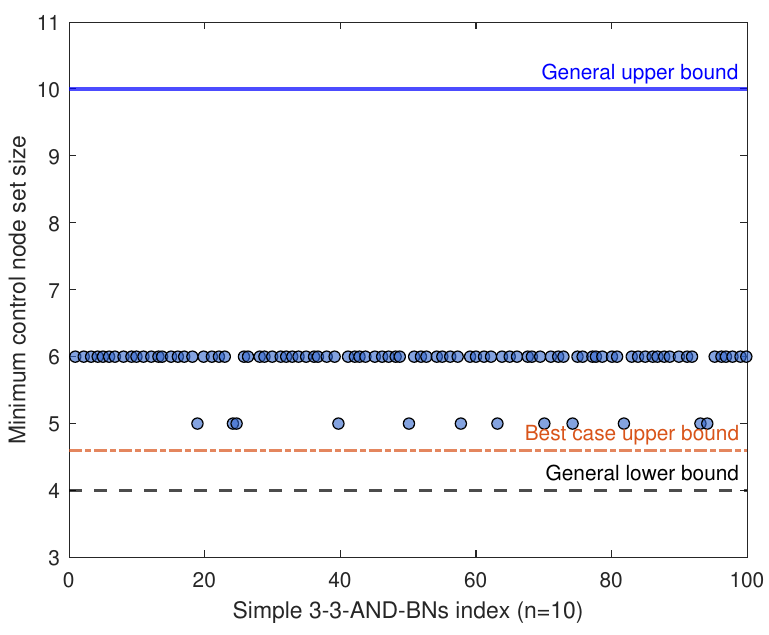}}
\hfil
\subfigure[]{\includegraphics[width=0.45\textwidth]{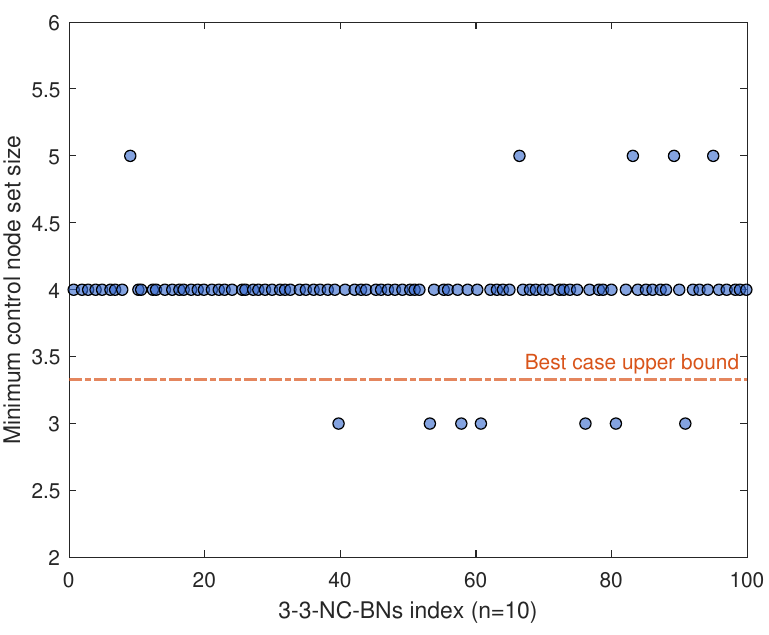}}
\hfil
\subfigure[]{\includegraphics[width=0.45\textwidth]{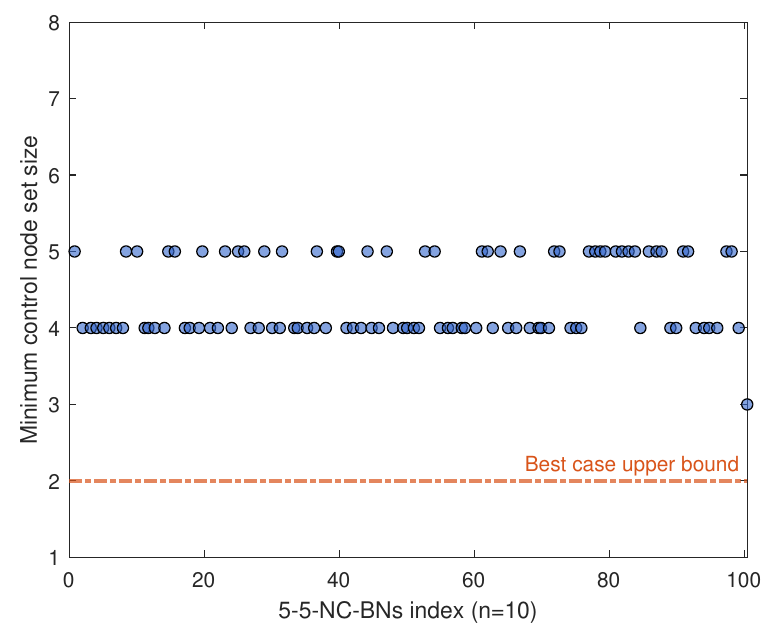}}
\caption{Comparison between the minimum control node set size and the derived bounds.}
\label{5}
\end{figure}

To evaluate the effectiveness of the derived bounds, we conduct three sets of experiments.
The first set includes 100 randomly generated $k$-$k$-XOR-BNs with parameters $(n=10, k=2)$ and
$(n=10, k=3)$.
The second set includes 100 randomly generated simple $k$-$k$-AND-BNs with parameters $(n=10, k=2)$ and
$(n=10, k=3)$.
The third set includes 100 randomly generated simple $k$-$k$-NC-BNs with parameters $(n=10, k=2)$ and $(n=10, k=3)$. For each generated BN, we perform a brute-force search to determine the minimum control node set that guarantees controllability. The results, obtained by MATLAB simulations on a MacBook Pro (Apple M4 Pro, 24 GB RAM), are summarized in Fig. \ref{5}, which shows that the size of the minimum control node set always lies between the derived lower and upper bounds.

In addition, we find that there exist some $k$-$k$-NC-BNs whose minimum control node sets are smaller than $\frac{n}{k}$, suggesting that the derived best case upper bound may be further improved. Nevertheless, it can still be observed that some $k$-$k$-NC-BNs are easier to control than any simple $k$-$k$-AND-BN.

\section{Conclusion}\label{section-6}
This paper comprehensively studied the minimum control node set problem for BNs with degree constraints. Specifically, by dividing nodes into three disjoint sets, extending the time to reach the target state, and utilizing necessary conditions for controllability, this paper derived the general lower bound, best case upper bound, worst case lower bound and general upper bound on the size of the control node set for four types of BNs ($k$-$k$-XOR-BNs, simple $k$-$k$-AND-BNs, $k$-$k$-AND-BN with negation and $k$-$k$-NC-BN), respectively.

In a word, the nontrivial lower and upper bound on the size of the minimum control node set were obtained, and it is worth mentioning that a number of meaningful results and phenomena were discovered based on the derived conclusions, as follows:
\begin{itemize}
  \item According to Theorem \ref{thm:xor-time}, when $t^{*}=2$, the general lower bound on the size of the control node set is $\frac{n}{2}$. On the other hand, the best case upper bound for 2-2-XOR-BNs is $\frac{n}{2}$ mentioned in Proposition \ref{prop:xor-lower}, which means that the general lower bound$=$best case upper bound in this case, i.e., these bounds are optimal.
  \item Note that Proposition \ref{proposition6} is a special case of Proposition \ref{proposition7}, that is, by letting $k=2$, the best case upper bound on the size of the control set of Proposition \ref{proposition7} becomes $\frac{1}{3}(n-1)+1$. On the other hand, Theorem \ref{thm:2-2-and} is also a special case of Theorem \ref{thm:k-k-AND}, that is, by letting $k=2$, the general lower bound on the size of the control set of Theorem \ref{thm:k-k-AND} becomes $\frac{1}{3}n$. Furthermore, $\frac{1}{3}(n-1)+1$ is close to $\frac{1}{3}n$ (resp., $\frac{k-1}{2k-1}(n-1)+1$ is close to $\frac{k-1}{2k-1}n$), indicating that the bounds given by Proposition \ref{proposition6} and Theorem \ref{thm:2-2-and} are almost optimal (resp., the bounds given by Proposition \ref{proposition7} and Theorem \ref{thm:k-k-AND} are almost optimal).
  \item According to Theorem \ref{thm:2-2-AND-negation}, the set of control node set is at least $\frac{1}{3}n$ when $|M_{1}|\leq\frac{2}{3}n$, which is consistent with the conclusion of Theorem \ref{thm:2-2-and} ($M_{1}=\emptyset$). In other words, Theorem \ref{thm:2-2-and} is a special case of Theorem \ref{thm:2-2-AND-negation}, which is also natural due to the fact that the simple 2-2-AND BN is a special case of the 2-2-AND-BN with negation. On the other hand, when $|M_{1}|>\frac{2}{3}n$, the larger the value of $|M_{1}|$, the larger the general lower bound given in Theorem \ref{thm:2-2-AND-negation}.
  \item More generally, according to Remark \ref{remark10},
%for each $l=l^{*}$,
%$\frac{k-(l^{*}+1)}{2k-(l^{*}+1)}n+\frac{1}{2k-(l^{*}+1)}\left(\sum_{j=1}^{(l^{*}+1)-2}j\cdot|M_{j}|\right)+\frac{(l^{*}+1)-1}{2k-(l^{*}+1)}\left(\sum_{j=(l^{*}+1)-1}^{\lfloor\frac{k}{2}\rfloor}|M_{j}|\right)>\frac{k-l^{*}}{2k-l^{*}}n+\frac{1}{2k-l^{*}}\left(\sum_{j=1}^{l^{*}-2}j\cdot|M_{j}|\right)+\frac{l^{*}-1}{2k-l^{*}}\left(\sum_{j=l^{*}-1}^{\lfloor\frac{k}{2}\rfloor}|M_{j}|\right)$,
%when $\sum_{j=1}^{l^{*}-1}j\cdot|M_{j}|+(2k-1)(|M_{l^{*}}|+\cdots+|M_{\lfloor\frac{k}{2}\rfloor}|)>kn$. Similarly, it can be seen that the larger $|M_{j}|,~j=1,\ldots,\lfloor\frac{k}{2}\rfloor$, the larger the general lower bound on the size of control node set, and the influence of the values of $|M_{1}|,\ldots,|M_{\lfloor\frac{k}{2}\rfloor}|$ on the value of the general lower bound is from small to large, that is, $|M_{\lfloor\frac{k}{2}\rfloor}|$ has the greatest impact on the value of the general lower bound.
  it can be seen that the larger of the values of $|M_{j}|,\; j=1,\ldots,\lfloor\frac{k}{2}\rfloor$, the larger the general lower bound given in Theorem \ref{thm:k-k-AND-negation}, and the influence of the values of $|M_{1}|,\ldots,|M_{\lfloor\frac{k}{2}\rfloor}|$ on the value of the general lower bound is from small to large, that is, $|M_{\lfloor\frac{k}{2}\rfloor}|$ has the greatest impact on the value of the general lower bound.
  \item Notably, when $k \geq 3$, the size of the minimum control node set $\frac{n}{k}$ in Proposition \ref{prop:nc} is less than $\frac{k-1}{2k-1}n$ in Theorem \ref{thm:k-k-AND}, which implies that some $k$-$k$-NC-BNs are easier to control than any $k$-$k$-AND-BN.
\end{itemize}
Notably, all of the above results involving the AND function also apply to the OR function.

However, there are still some open issues that need to be resolved in the future. First, the gap between the bounds given by Proposition \ref{prop:xor-lower} and Theorem \ref{thm:4-4-xor} is significant, so it is necessary to narrow this gap, and some improvements on XOR-BNs \cite{fok25} were developed based on this work after its submission. Moreover, a general upper bound or a worst case lower bound
for $k$-$k$-AND-BN with negation (including the case of $k=2$) is not given in the paper, so it is necessary to derive these bounds further.

\end{document}